\documentclass[graybox,natbib,nosecnum]{svmult}
\bibpunct{(}{)}{;}{a}{}{,} % suppress commas between author-names and year

\pdfoutput=1   %forces use of pdflatex. Disable if you prefer to use .eps or .ps figures.
\usepackage{amsmath}
\usepackage{mathptmx}       % selects Times Roman as basic font
\usepackage{helvet}         % selects Helvetica as sans-serif font
\usepackage{courier}        % selects Courier as typewriter font
\usepackage{type1cm}        % activate if the above 3 fonts are
\usepackage{makeidx}         % allows index generation
\usepackage{graphicx}        % standard LaTeX graphics tool
\usepackage{multicol}        % used for the two-column index
\usepackage[bottom]{footmisc}% places footnotes at page bottom
\usepackage[normalem]{ulem}	% for strike-through of text with \sout{}  
\usepackage{hyperref}  %for hyperlinks
\usepackage{url}  %for hyperlinks

\usepackage{soul}   % for high-lighting of text
\makeindex             % used for the subject index
\begin{document}

% ------------------------------------------------------------------------
% New commands
%
\def\ltsima{$\; \buildrel < \over \sim \;$}
\def\lsim{\lower.5ex\hbox{\ltsima}}
\def\gtsima{$\; \buildrel > \over \sim \;$}
\def\gsim{\lower.5ex\hbox{\gtsima}}
% -------------------------------------------------------------------------

\title*{Planet Occurrence: Doppler and Transit Surveys}
\titlerunning{Doppler and Transit Surveys}

\author{Joshua N.\ Winn\inst{a} and
Erik Petigura\inst{b}}
\authorrunning{Winn and Petigura}
% First names are abbreviated in the running head.
% If there are more than two authors, 'et al.' is used.
%
\institute{$^a$Princeton University, Princeton, NJ, USA \email{jnwinn@princeton.edu},\\
$^b$University of California, Los Angeles, CA, USA \email{petigura@astro.ucla.edu}}

%\and
%

\maketitle

\abstract{Prior to the 1990s, speculations about the occurrence of
  planets around other stars were based only on planet formation
  theory, observations of circumstellar disks, and the knowledge that
  at least one seemingly ordinary star is the host
  of four terrestrial planets, two gas giants, and two ice giants.
  Since then, Doppler and transit surveys have
  been exploring the population of planets around other Sun-like stars,
  especially those with orbital periods shorter than a few years.
  Over the last decade, these surveys have risen to new heights with
  Doppler spectrographs with a precision better than 1~m~s$^{-1}$ precision, and space
  telescopes capable of detecting the transits of Earth-sized planets.
  This article is a brief introductory review of the knowledge of
  planet occurrence that has been gained from these surveys.}
\section{Introduction}

If, in some cataclysm, all our knowledge about
exoplanets from Doppler and transit surveys
were to be destroyed, and only one brief sentence
passed on to the next generation of
astronomers, what statement would contain the most helpful
information?\footnote{Adapted from book I, chapter 1, verse 2 of
\citet{Feynman1963}.}
Here is one possibility:\\

\noindent {\it  At least a third of Sun-like stars have several Earth-to-Neptune-sized planets -- and a tenth have
  giant planets -- orbiting between 0.05 and 1~AU.}\\
  
\noindent If we could preserve a
mathematical function instead of a
sentence, we might choose 
\begin{equation}
\label{eq:period-distribution}
\frac{dn}{d\log P} \approx C\,P^\beta\,\left[1 - e^{-(P/P_0)^{\gamma}} \right],
\end{equation}
along with the values of the constants $C$, $\beta$, $\gamma$, and $P_0$ that apply to planets of different sizes \citep{Howard+2012}.
When integrated from $\log P_1$ to $\log P_2$, this function
gives the average number of planets per star
having orbital periods between $P_1$ and $P_2$.
The function is an example of an {\it occurrence rate density},
in which the average number of planets
per star ($n$, the {\it occurrence rate}) is differentiated with
respect to chosen
characteristics of the system (making it a {\it density}).

Even more helpful to astronomers starting from scratch would be a computer program that produces
random realizations of the key properties of planetary systems that are statistically
consistent with everything we have learned from our surveys \citep[see, e.g.,][]{He+2019}.
Ideally, such a ``generative model'' would include planetary properties besides $P$ such
as planetary mass and orbital
eccentricity, as well as stellar properties like mass, metallicity, and age. The model would also take into account that the occurrence of one type of planet depends on the properties of the other planets in the system, i.e., it would incorporate {\it conditional} occurrence rates.

Transmitted in any of these forms to our descendants, the occurrence information would dispel any prejudice that all planetary systems should
resemble the Solar System,
help them design their instruments to detect exoplanets,
and inspire their theories for planet formation. However, the real point of this thought experiment 
is to convey
that {\it occurrence} is a topic of central importance in exoplanetary science
and that it can be treated at many levels of sophistication.
The subject of this introductory review is the knowledge we have gained
about planet occurrence from Doppler and transit surveys. The details of the
Doppler and transit techniques themselves
are left for other reviews,
such as those by \citet{LovisFischer2010}, \citet{Winn2010}, and
\cite{Wright2018}.  Here, we will simply remind ourselves of the
key properties of the Doppler and transit signals:
\begin{eqnarray}
  K & = & \frac{0.64~{\rm m~s}^{-1}}{\sqrt{1-e^2}} \left( \frac{P}{1~{\rm day}} \right)^{\!\!-1/3}
  \frac{(M/M_\oplus)\sin I}{(M_\star/M_\odot)^{2/3}}, \\
  \delta & \approx & 8.4\times 10^{-5}
   \left( \frac{R/R_\star}{R_\oplus/R_\odot} \right)^{\!\!2},~~p_{\rm tra} = \frac{0.0046}{1-e^2}~\left(\frac{R_\star/a}{R_\odot/1~{\rm AU}}\right) 
\end{eqnarray}
where $K$ is the radial-velocity semi-amplitude; $\delta$ is
the fractional loss of light during transits; $p_{\rm tra}$ is the
probability for a randomly-oriented orbit to exhibit transits; $a$,
$P$, $e$, and $I$ are the orbital semi-major axis, period,
eccentricity, and inclination; $M$ and $R$ are the mass and radius of the planet,
and $M_\star$ and $R_\star$ are those of the star.

The next section describes methods for occurrence calculations and is
followed by two sections on the results.
Because the surveys have revealed major differences in occurrence between
giant planets and small planets, with a dividing line just above
4~$R_\oplus$ or 20~$M_\oplus$, the results for small planets and giant planets
are presented separately. After that comes a review of what
is known about other types of stars, and a discussion of
future prospects.

\section{Methods}
\label{sec:methods}

Life would be simple if planets came in only one type and we could
detect them unerringly.  We would search $N$ stars, detect $N_{\rm
  det}$ planets, and conclude that the occurrence rate
  is $n \approx N_{\rm det}/N$, with an uncertainty dictated
  by counting statistics.  Unfortunately,
detection is not assured: small signals can be lost in the noise.
If the detection probability $p_{\rm det}$ were the same
for every star that was searched, then effectively
we would only have searched $N_{\rm eff} = p_{\rm det}N$ stars, and the estimated occurrence rate would be $n \approx N_{\rm det}/(p_{\rm det}N)$.

\begin{figure}
\includegraphics[scale=0.6]{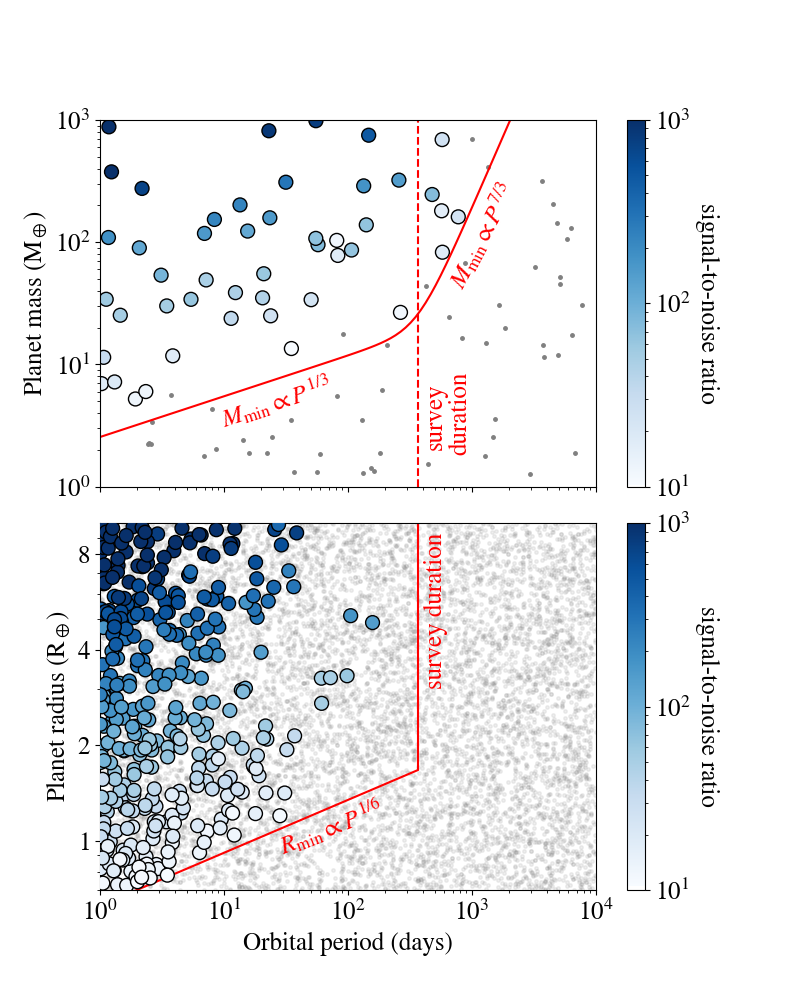}
\caption{
%\vskip -0.2in
{\bf Idealized Doppler and transit surveys}.
The top panel shows the results
of a Doppler survey
in which 100 Sun-like stars 
are each observed 50 times over one year with 1~m~s$^{-1}$ precision.
The bottom panel shows the
results of a transit survey in which $10^4$ stars
are observed continuously for a year with a
photometric precision corresponding to $3\times 10^{-5}$ per 6 hours of data.
Each star is assumed to have one planet on a randomly-oriented circular
orbit, with random properties drawn from log-uniform distributions between the plotted limits.
Colored points are planets
detected with a signal-to-noise ratio (SNR) exceeding 10. Gray points
are undetected planets.
In the Doppler survey,
the threshold mass
scales as $P^{1/3}$ for
periods shorter than the survey duration.
For longer periods, the threshold mass
increases more rapidly, with an exponent depending on the desired
false-alarm probability \citep{Cumming2004}.
The transit survey includes more stars, because they can be monitored
simultaneously, but 
detects a smaller
fraction of the planets and is more strongly biased toward short periods.
For periods shorter than survey duration,
the threshold radius varies as
$P^{1/6}$ \citep{Pepper+2003}.
For longer periods it is impossible to observe more than one transit, making any
detections more ambiguous.
\label{fig:survey}
}
\end{figure}

In reality, $p_{\rm det}$ depends strongly on the characteristics of the star and planet (see Figure~\ref{fig:survey}).
Detection is easier for brighter stars, shorter
orbital periods, and larger and more massive planets.  For this
reason we need to group the planets according to orbital period and other salient characteristics for detection: the radius $R$ for transit surveys
and the minimum mass $M$ for Doppler surveys.
Actually, for Doppler surveys the observable quantity is $M\sin I$ rather than
$M$, but this is a minor issue for occurrence studies
if we are willing to assume that planetary systems are
oriented randomly relative to our line of sight,
implying $\langle \sin I\rangle = \pi/4$.
Transit surveys have a more serious complication: most planets
produce no signals at all, because $I$ needs to be very
close to $90^\circ$ for transits
to occur. With this consideration in mind,
our estimates for the occurrence rate become
\begin{equation}
\label{eq:occurrence}
  n_i \approx \frac{N_{{\rm det},i}} {N_{{\rm eff},i}}~{\rm where}~N_{{\rm eff},i} =
  \begin{cases}
  \sum_{j=1}^N p_{{\rm det},ij} & \text{for Doppler surveys, and} \\
  \sum_{j=1}^N p_{{\rm det},ij} \times p_{{\rm tra},ij} & \text{for transit surveys,}
\end{cases}
\end{equation}
where the index $i$ refers to a group of planets sharing similar
characteristics,
and the index $j$ specifies the star that was searched.

This conceptually simple method has been the basis of many
investigations. The results of Doppler surveys are sometimes presented as a
matrix of occurrence rates for rectangular bins in the space of
$\log M$ and $\log P$ \citep[e.g., Figure~1 of ][]{Howard+2010}.
For transit surveys, the bins are in the
space of $\log R$ and $\log P$ (e.g., Figure 4 of \citealt{Howard+2012}).
Ideally, the bins should be large enough to contain
many detected planets, and yet small enough that the
occurrence rate density and the effective number of searchable
stars do not vary too much across the bin.
When the number of detected planets in each bin is
modeled as a Poisson random variable, the maximum-likelihood estimate of the occurrence
rate in each bin is
$n_i = N_{{\rm det},i} / N_{{\rm eff},i}$ \citep{TabachnikTremaine2002}.
More careful handling 
is needed when there are substantial uncertainties
in $N_{{\rm det},i}$ arising from uncertainties in the relevant
planetary and stellar characteristics \citep{ForemanMackey+2014}.

Another approach is to posit a parameterized
function for the occurrence rate density,
such as Equation~\ref{eq:period-distribution}, but often in two or more
dimensions. For example, Doppler surveyors have
often chosen a double power law in mass and period:
\begin{equation}
\label{eqn:powerlaw}
\frac{\partial^2n}{\partial\log M~\partial\log P} = C\,M^\alpha P^\beta,
\end{equation}
where $C$, $\alpha$, and $\beta$ are constants that are
estimated
by maximizing the likelihood
of drawing planets from the distribution
that match the number and characteristics of the
detected planets.
For details on this approach, see \citet{TabachnikTremaine2002},
\citet{Cumming+2008}, and \citet{Youdin2011}.
Subsequent authors have increased the
level of sophistication
by applying methods from Bayesian hierarchical inference
\citep{ForemanMackey+2014},
and ``likelihood-free'' approximate Bayesian computation \citep{Hsu+2018}.

Most studies report the occurrence rate density as a function of the properties
of a planet, regardless of any other planets in the system.
More difficult is quantifying the {\it multiplicity} of planetary
systems, the number of planets that orbit together around the same
star.  For Doppler surveys, one problem is that the star is pulled by
all the planets simultaneously.  As a result, the detection probability for a
given planet depends on the properties of any other planets ---
especially their periods --- and on the time sampling and total timespan of the Doppler observations.
For transit surveys, the detectability of one planet is nearly independent of any others because transits only rarely overlap \citep{Zink+2019}.
Instead, the main problem is a degeneracy
between multiplicity and the mutual inclinations between orbits.
A star with only one detected transiting planet
might lack additional planets, or it might have several
planets only one of which happens to transit. This
degeneracy can be broken -- with difficulty -- by
modeling transit durations
\citep{He+2019} or
transit-timing variations \citep{Zhu+2018},
or by combining the results of a transit survey
with a Doppler survey of similar stars \citep{TremaineDong2012, Figueira+2012}.

Doppler surveys have discovered on the order of $10^3$ planets.
Among the most informative surveys for planet occurrence
were based on observations
with the High Resolution Echelle Spectrometer (HIRES) on the Keck~I
10-meter telescope \citep{Cumming+2008,Howard+2010} and the High
Accuracy Radial-velocity Planet Searcher (HARPS) on the La Silla
3.6-meter telescope
\citep{Mayor+2011,Fernandes+2019}. Each instrument was used to
monitor $\approx$500 stars for about a decade, with a precision of a
few meters per second. Additional information comes from a few
lower-precision and longer-duration surveys \citep[see,
  e.g.,][]{LovisFischer2010}.
Another valuable resource
is the California Legacy Survey
\citep[CLS;][]{Rosenthal+2021},
a meta-survey of 718 stars
based on $\sim$$10^5$ archival
Doppler measurements spanning
several decades.

For transits, surveys with
ground-based telescopes
have discovered several hundred
planets, but they are not often used
for occurrence studies because
the limitations of ground-based data
make it difficult to characterize the sample of searchable stars and
calculate the detection probabilities.
Instead, our most important resources
are the NASA space-based surveys Kepler, K2, and the
Transiting Exoplanet Survey Satellite (TESS).
Kepler used a 1-meter space telescope to
measure the brightness of about 150{,}000 stars from 2009 to 2013
\citep{Borucki2016}. The typical photometric precision over a 6-hour
time interval was on the order of $10^{-4}$,
which was sufficient to detect
about 4{,}000 planets \citep{Lissauer+2023}.
K2 used the same telescope to survey 19 fields along the
ecliptic in 80-day increments from 2014 to 2018 \citep{Howell+2014}. Since 2018, TESS has been surveying the entire sky in month-long campaigns with four 0.1-meter cameras \citep{Ricker+2015}.
Although
K2 and TESS data have been used to
discover several thousand planets and planet candidates, the
Kepler mission was more sensitive to planets with longer
periods and smaller sizes. Therefore, our knowledge of transiting planet demographics is still anchored by Kepler, with K2 and TESS providing supplementary information by searching stars with a wider variety of masses, ages, and locations within the galaxy.

\section{Giant planets}
\label{sec:giant}

\begin{figure}
\includegraphics[scale=0.315,angle=0]{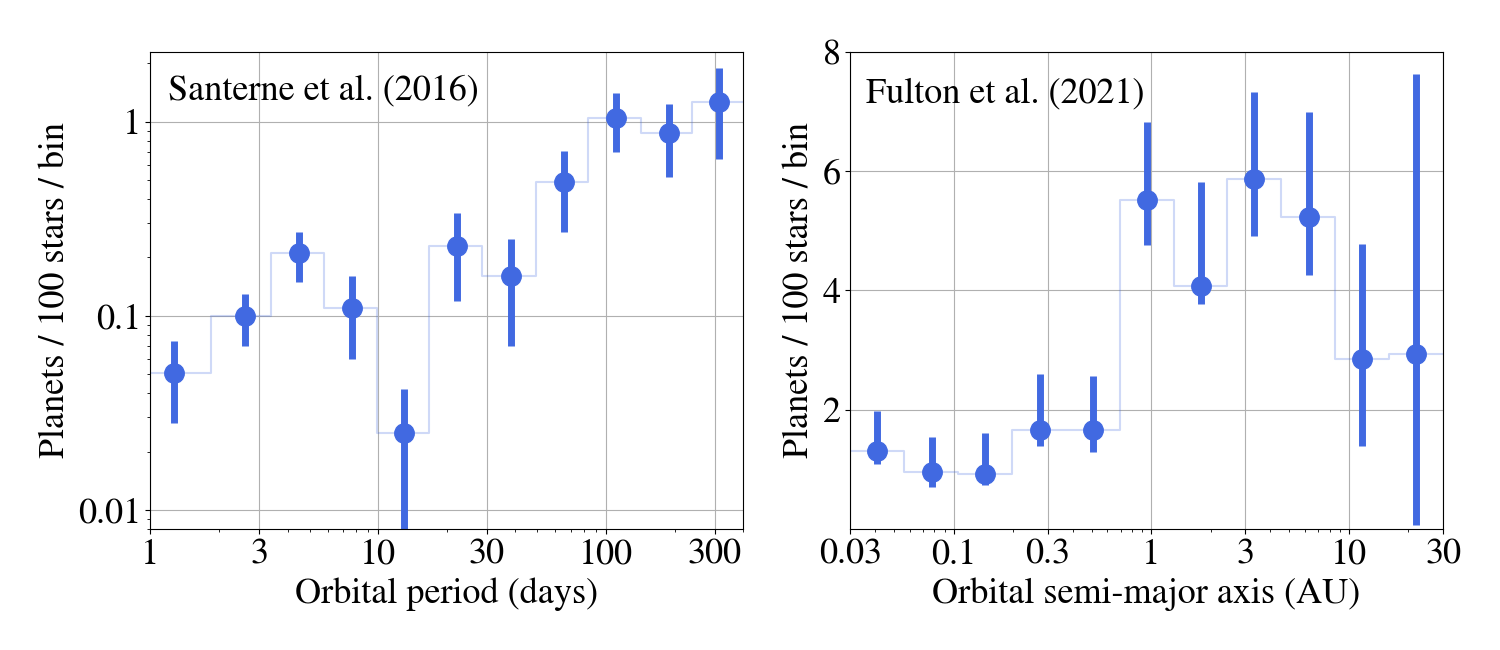}
\caption{
Occurrence rate density
of giant planets, expressed as the 
average number of planets per 100 stars
within each bin of either period
(left) or semi-major
axis (right). The data at left were derived
from the Kepler transit survey
by \citet{Santerne+2016}.
The data at right were derived
from the Doppler-based
California Legacy Survey by \citet{Fulton+2021}.
\label{fig:giants}}       % Give a unique label
\end{figure}

\runinhead{Overall occurrence} For giant planets,
some key references
are \citet{Fulton+2021}, who computed planet occurrence
based on the CLS,
\citet{Wittenmyer+2020}, who did the same
for the 18-year Anglo-Australian Planet Search,
\citet{Santerne+2016} and \citet{Petigura+2018}, who used the Kepler
transit survey,
and 
\citet{Fernandes+2019},
who combined the Doppler results of \cite{Mayor+2011} and the  Kepler results of \citet{Santerne+2016}.
Although these studies differ in detail,
their overall message is that
approximately one-tenth of
Sun-like stars
have giant planets
with orbital
distances
smaller than 1\,AU.
For orbital distances from 1--10\,AU,
the fraction rises to about one-third,
and there is tentative evidence
that $dn/d\log a$ has a broad
peak centered at about
3\,AU (see Figure~\ref{fig:giants}).
This peak
might be related to the location of the
``snow line'' in protoplanetary disks, which plays an
important role in the
theory of giant-planet formation via core accretion;
beyond this line it is cold enough for water to freeze,
increasing the mass of solid material
that is available to help a growing planet achieve the critical mass for runaway gas accretion \citep{Pollack+1996,Lecar+2006}.
For periods shorter than a few years, the distribution of planet
masses, $dn/d\log M$, appears
to be roughly constant 
between about 30 and 1{,}000\,$M_\oplus$
\citep{Fernandes+2019, Fulton+2021}.

\runinhead{Metallicity} The earliest Doppler surveys found
the occurrence of giant planets 
to be associated with a high heavy-element abundance
of the host star \citep{Gonzalez+1997}.
\citet{FischerValenti2005}
found $dn/dZ \propto Z^2$,
where $Z$ is the mass fraction
of ``metals'' (elements heavier than
helium).
Another valid description
of the data is that giant-planet
occurence is low and independent
of metallicity for $Z\lsim Z_\odot$, and rises steeply with metallicity for $Z\gsim Z_\odot$
\citep{Santos+2003, Fulton+2021}.

This ``metallicity effect'' is widely
interpreted as supporting evidence for the core accretion theory of giant planet formation.
The logic is that
the rapid assembly of a massive solid core --- an essential step in
the theory --- is easier to achieve in a metal-rich protoplanetary
disk.
\citet{Santos+2017}
found the occurrence of
companions more massive than 4~$M_{\rm Jup}$ to
be independent of metallicity and suggested that such objects
form by gravitational instability
rather than core accretion.  \citet{Schlaufman2018} reached
a similar conclusion and went so far as to say that companions more
massive than 10~$M_{\rm Jup}$ should not be considered planets. 
Complicating the interpretation,
\citet{Buchhave+2018}
found that giant planets with $a \gsim 2$\,AU and low eccentricities
are not preferentially
found around high-metallicity stars,
suggesting that the
metallicity effect is related
to giant-planet {\it migration} rather than
{\it formation}.

\runinhead{Hot Jupiters} Easy to detect, but intrinsically rare,
hot Jupiters have an occurrence rate of 0.5--1\% for periods between 1
and 10~days. They are even rarer for periods shorter than one day
\citep{Howard+2012,SanchisOjeda+2014}.
There is a 3$\sigma$
discrepancy between the rate of 0.8--1.2\% measured in Doppler surveys
\citep{Wright+2012,Mayor+2011} and 0.6\% measured using Kepler
data \citep{Howard+2012,Petigura+2018}. This is despite the similar
metallicity distributions of the stars that were searched
\citep{Guo+2017}. While we should never lose too much sleep over
3$\sigma$ discrepancies,
an interesting explanation was
offered by \cite{MoeKratter2021}: giant planet
formation is suppressed in
binary star systems with separations
less than about 10\,AU.
Transit surveys include
many such close binaries
in their search samples \citep{Bouma+2018},
but Doppler surveys
generally exclude them,
which could
boost the inferred occurrence of hot Jupiters in Doppler
surveys by a factor of 1.5--2.

\runinhead{Conditional occurrence rates}
Given the existence of a close-orbiting
giant planet, 
what is the chance of finding another
planet around the
same star?
Many authors have attempted to measure
such conditional occurrence rates
because they might provide clues about the
formation and evolution of giant planets.
For example,
\citet{Huang+2016} used Kepler data
to show that hot Jupiters ($P=$1--10\,days) are ``lonelier'' than warm Jupiters (10--100\,days), in that
they have a lower occurrence rate
of companions with periods
shorter than 50 days and radii larger
than 2\,$R_\oplus$. However,
hot and warm Jupiters have similar
occurrence rates of more distant
giant planets \citep{SchlaufmanWinn2016,Bryan+2016,ZinkHoward2023}.
These results suggest that
the formation of a hot Jupiter involves events
that destroy or suppress
the formation of any
other planets within $\approx$0.5\,AU,
as expected in the
theory of high-eccentricity
migration \citep{RasioFord1996}.

\runinhead{Other properties} The giant-planet population is
distinguished by other features. Their orbital eccentricities range from zero to nearly unity with a mean value of about 0.3 \citep{Kipping+2013}. Their occurrence
seems to fall precipitously for masses above $\approx$10~$M_{\rm
  Jup}$, at least for orbital distances smaller than a few AU.
Because of this low occurrence, the mass range from 10--80\,$M_{\rm
  Jup}$ is often called the ``brown dwarf desert''  \citep{GretherLineweaver2006,Sahlmann+2011,Santerne+2016,Triaud+2017}.
  %As mentioned
%earlier, the inhabitants of this desert are not associated
%with high-metallicity stars, making them unlike Jovian-mass
%planets\citep{Santos+2017,Schlaufman2018}.
Occasionally, we find two
giant planets in a mean-motion resonance \citep{Wright+2011}.
The rotation of the star can be grossly misaligned with the orbit of the planet, especially if the star is more massive than about
1.2~$M_\odot$ \citep{Albrecht+2022}.

\section{Smaller planets}
\label{sec:small}

\runinhead{Overall occurrence} At least a third of Sun-like stars have ``miniature Solar Systems'' consisting of several
planets with periods shorter than a year and
sizes in between those of Earth and Neptune.
Planet formation theories
generally did not predict this profusion of close-orbiting planets.
Indeed, some of the most detailed theories predicted that such planets would be especially rare
\citep{IdaLin2008}. Their surprisingly high abundance led to new
theories in which small planets can form in short-period orbits,
rather than forming farther away from the star and migrating
inward \citep[see, e.g.,][]{HansenMurray2012,ChiangLaughlin2013}.

Doppler surveys provided our first glimpse at this population of
planets. For periods shorter than 50 days and masses between 3
and 30\,$M_\oplus$, two independent Doppler surveys found the
occurrence rate to be $(15\pm 5)\%$ \citep{Howard+2010} and $(27\pm
5)\%$ \citep{Mayor+2011}.
Soon after, the Kepler mission
revealed this
population in more vivid detail \citep{Howard+2012}.
For example, \cite{Zhu+2018} used Kepler data to show
that $(30\pm 3)\%$ of Sun-like stars
harbor several planets with
periods shorter than 400 days and
sizes between 1 and 4\,$R_{\oplus}$.
Any association between their occurrence
and the metallicity of the host star
is weaker
than for giant planets
\citep{Buchhave+2012,Mulders+2016,Winn+2017,Wilson+2018,Petigura+2018}.

\begin{figure}[ht!]
\includegraphics[scale=0.315,angle=0]{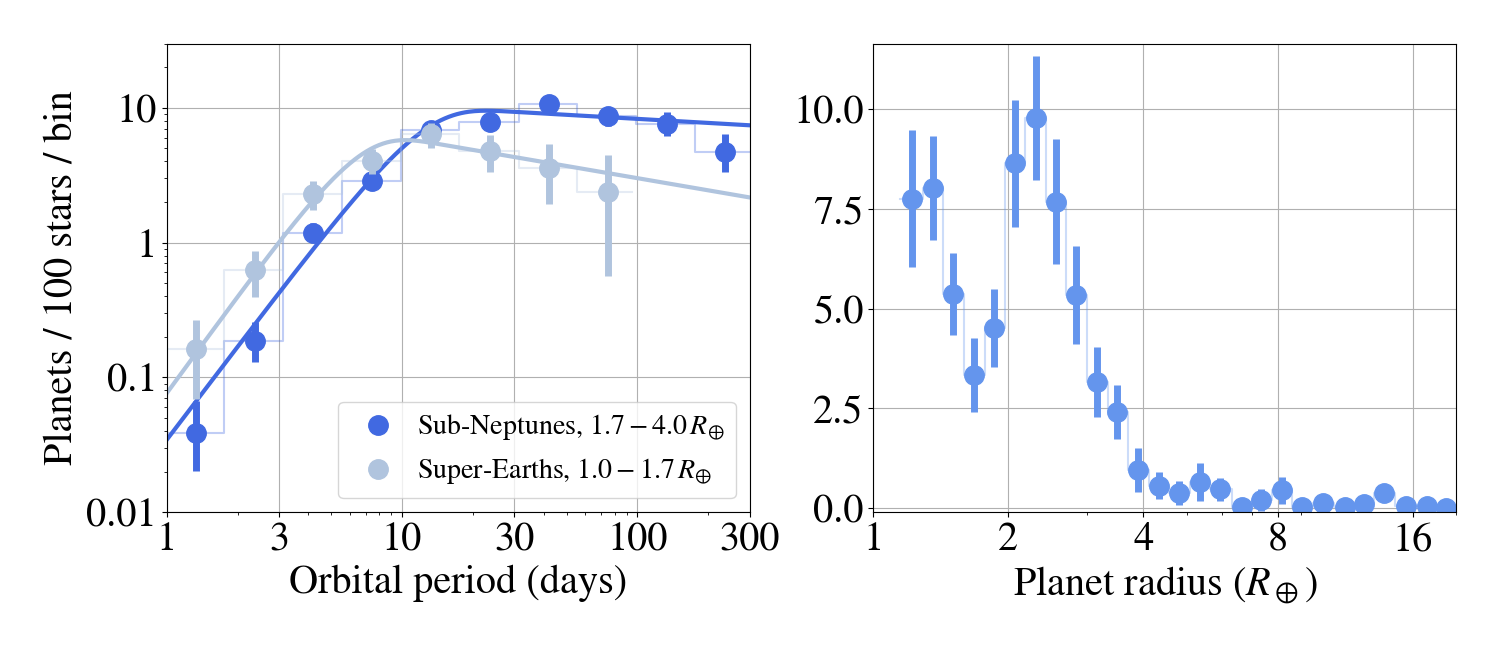}
\caption{Occurrence rate density
of small planets, expressed as the 
average number of planets per 100 stars
within each bin of either period (left) or
radius (right). The period distributions are from \citet{Petigura+2018},
with best-fit functions of the
form given by Equation~\ref{eq:period-distribution}.
The radius distribution is from \citet{Fulton+2021}
and refers to planets
with periods shorter than 100 days.
The dip at 1.7\,$R_\oplus$ appears to separate
solid ``super-Earths'' from gas-sheathed ``sub-Neptunes''.
\label{fig:smallplanets}} 
\end{figure}

\runinhead{Size and period}
Within the innermost AU of planetary
systems, planets with sizes between
1 and 4\,$R_\oplus$
are about an order of magnitude
more abundant than planets with sizes
between 4 and 16\,$R_\oplus$.
The occurrence rate density
$dn/d\log P$
of 1-4\,$R_\oplus$ planets
rises with period out to about 10 days
and is nearly flat between 10--400 days.
The left panel of Figure~\ref{fig:smallplanets} shows
the occurrence rate densities derived by \citet{Petigura+2018}
using  Kepler data,
for two different size categories.

\runinhead{Orbital spacings} As noted above, small planets frequently occur in closely-spaced systems.
\cite{Zhu+2018} found the typical system to consist
of three planets having periods shorter than 400 days.
The ratios of orbital periods
between adjacent planets tend to be in the
range between 1.5 and 5 \citep{Fabrycky+2014}.
With reference to the mutual Hill radius
(relevant to dynamical stability),
\begin{equation}
  a_{\rm H} \equiv \left( \frac{M_{\rm in} + M_{\rm out}}{3M_\star} \right)^{\!1/3}
  \left( \frac{a_{\rm in} + a_{\rm out}}{2} \right),
\end{equation}
the typical spacing is
10--30~$a_{\rm H}$ \citep{FangMargot2013}.  At the lower end of this distribution,
the systems flirt with instability \citep{Deck+2012,PuWu2015}.
A few percent of
the Kepler systems are in or near mean-motion resonances,
suggesting that the orbits have been sculpted by planet-disk gravitational interactions. Resonant and near-resonant systems offer the gift of
transit-timing variations (TTVs), the observable manifestations of planet-planet gravitational interactions.
In some cases, modeling the TTVs leads to precise constraints on planet masses, eccentricities, and inclinations (see, e.g., \citealt{Carter+2012} and \citealt{Agol+2021}), although
there are often degeneracies between these quantities \citep{Lithwick+2012}. Analyses of TTVs, and other lines of evidence, have shown that the compact multiple-planet systems tend to have orbits that are nearly circular
\citep{HaddenLithwick2014,Xie+2016,VanEylenAlbrecht2015} and coplanar
\citep{Fabrycky+2014}.

\runinhead{Radius gap}
For planets with periods shorter than 100 days,
the occurrence rate density $dn/d\log R$ shows a dip
centered at $R\approx 1.7_\oplus$ \citep[][see
the right panel of Figure~\ref{fig:smallplanets}]{Fulton+2017,VanEylen+2017}.
The location of this dip is often used as the dividing line
between ``super-Earths'' and ``sub-Neptunes''.
Super-Earths tend to have overall densities consistent
with rocky planets. Sub-Neptunes tend to have lower
densities, suggestive of a rocky planet with an outer layer of
hydrogen-helium gas constituting a few percent of
the total mass. Possibly, super-Earths
are sub-Neptunes that lost their atmospheres
due to the host star's high-energy radiation \citep{OwenWu2013,
  LopezFortney2013} or the gradual leakage of the
  rocky planet's heat of formation \citep{Ginzburg+2018}.

\runinhead{Hot Neptunes} Both Doppler and transit
surveys found a very low occurrence rate for planets
with periods shorter than a few days and
sizes between 2 and 6~$R_\oplus$ or
masses between 10 and 100~$M_\oplus$
\citep{SzaboKiss2011,Mazeh+2016}.
This ``hot Neptune desert'' may be another
consequence of atmospheric erosion.
Those few hot Neptunes that do exist are strongly
associated with metal-rich stars 
and tend to have planetary companions
in closely-spaced coplanar orbits, 
making them similar to giant planets and unlike smaller planets \citep{Dong+2018}.

\runinhead{Conditional occurrence rates}
Two super-Earths or two sub-Neptunes orbiting the same star tend to have more similar sizes than two planets of the same category
drawn from the entire collection of planetary systems. Their similar sizes
(and regular spacings) cause planets within a given system to resemble ``peas in a pod'' \citep{Weiss+2018, MillhollandWinn2021}. It seems logical that
planets forming in a similar environment should resemble
each other; for more on this phenomenon and
its interpretation, see \citet{Weiss+2023}.

Another interesting conditional occurrence rate is that
of distant giant planets around stars that harbor
short-period super-Earths and sub-Neptunes.
By combining the results of Doppler and transit surveys, \cite{ZhuWu2018} found that the
existence of a compact inner system boosts the
odds of finding a giant planet with a period of a few years from about 10\% to 30\%. Similar results
were obtained by \citet{Bryan+2019}.
Complicating the situation,
\citet{Rosenthal+2022} found evidence for a weaker boost
and \citet{Bonomo+2023}
found no evidence for any boost, seemingly contradicting
the earlier results. A possible resolution is that the
boost is specific to metal-rich stars \citep{Zhu2023}.

\runinhead{Earth-like planets} A goal with broad appeal is measuring
the occurrence rate of Earth-sized planets orbiting Sun-like stars
within the ``habitable zone'', the range of distances from the star
where the surface temperature of a rocky planet
could plausibly allow for oceans of liquid water. The 
  Kepler mission provided the best available data for this purpose.
However, even Kepler was barely sensitive to such planets. The
number of detections was on the order of 1--10,
depending on the
definitions of ``Earth-sized'', ``Sun-like'' and ``habitable
zone''. The desired quantity can be written
\begin{equation}
  \eta_\oplus \equiv \int_{R_{\rm min}}^{R_{\rm max}} \int_{S_{\rm min}}^{S_{\rm max}}
\frac{\partial^2n}{\partial\log S~\partial\log R}~d\log S~d\log R, 
\end{equation}
where $S$ is the flux
the planet receives from the star, and the integration limits are chosen to select planets likely to have a solid surface with a temperature and pressure suitable for liquid water \citep{Kasting+2014, Kopparapu+2014}.

The Kepler team published a series of papers reporting steady advances in quantifying the efficiency of planet detection, eliminating false positives,
understanding instrumental artifacts, and improving the characterization of the
stars that were searched.
The most recent study, by \cite{Bryson+2021},
summarized previous work on this topic and
found $\eta_\oplus$ to be between $0.37^{+0.48}_{-0.21}$
and $0.88^{+1.28}_{-0.51}$.
These two estimates are based on
different assumptions made
in extrapolating the occurrence rate from
larger planets and shorter periods.
\clearpage

\vspace*{1pt}
\section{Other types of stars}
\label{sec:other}

Almost all of the preceding results pertain to main-sequence stars with
masses between 0.5 and 1.2~$M_\odot$, i.e., spectral types from late K to late F.
Stars with masses between 0.1 and 0.5~$M_\odot$, the M
dwarfs, are not as thoroughly explored, especially near the low end of
the mass range. However these stars are very attractive for planet
surveys. Their small masses and sizes lead to larger Doppler
and transit signals, all other
things being equal.
Furthermore, planets in the habitable zones of M dwarfs have
short orbital periods that make
transits more likely and are
convenient for observers \citep{Gould+2003, NutzmanCharbonneau2008}.

The occurrence rate of giant planets with $a \lsim 1$\,AU
is lower around M dwarfs
than FGK dwarfs by at least a factor
of a few \citep{Endl+2006, Cumming+2008, Johnson+2010b, Bonfils+2013,Bryant+2023, Gan+2023}.
On the other hand, super-Earths
and sub-Neptunes with $a \lsim 1$\,AU
are several times {\it more} common around M dwarfs
than FGK dwarfs
\citep{Howard+2012,Mulders+2015, DressingCharbonneau2015}.
An implication is that the nearest habitable-zone planets
are probably around M dwarfs, and indeed,
Doppler surveys have turned up two
potentially habitable planets around very nearby
M dwarfs: Proxima~Cen \citep[1.3~pc,][]{AngladaEscude+2016} and
Ross~128 \citep[also known as Proxima~Vir; 3.4~pc,][]{Bonfils+2018}.
As is the case for FGK dwarfs, the small planets
around M dwarfs are often organized into compact systems
of multiple planets
\citep{Muirhead+2015,Gaidos+2016,BallardJohnson2016}.
There is also evidence that the planet population around M dwarfs exhibit both the hot Neptune
desert and the radius gap \citep{Hirano+2017},
although the nature of the planets above and below the radius gap is debated.
\cite{LuquePalle2022} argued that the sub-Neptunes
around M dwarfs are not gas-ensheathed rocky
planets, but are instead ``water worlds'' with a high
abundance of volatile elements, while \cite{Rogers+2023} argued that there is not yet conclusive evidence for a population of water worlds.

Beyond the scope of this review, but nevertheless fascinating, are the
occurrence rates that have been measured in Doppler and transit
surveys of other types of stars:
A stars \citep{Zhou+2019, Beleznay+2022},
evolved stars
\citep{Johnson+2010,Reffert+2015},
stars in open clusters
\citep{Mann+2017,Christiansen+2023} and globular clusters
\citep{Gilliland+2000,MasudaWinn2017},
binary stars \citep{Armstrong+2014},
brown dwarfs \citep{He+2017}, 
thick-disk and halo stars \citep{Zink+2023},
and white dwarfs \citep{Fulton+2014,VanSluijsVanEylen2018}.
Even neutron stars have been surveyed, using the Doppler-like
technique of pulsar timing
\citep{Kerr+2015, Behrens+2020, Nictu+2022}.

\section{Future Prospects}
\label{sec:future}

Improving upon the state of the art in Doppler
and transit surveys will not be easy, but
efforts are underway.
Surveys of M dwarfs are being conducted
with stabilized high-resolution infrared spectrographs,
a relatively new technological development
\citep[see, e.g.,][]{Mahadevan+2012,Sabotta+2021}.
With a new generation of optical Doppler spectrographs,
such as MAROON-X \citep{Seifahrt+2016},
ESPRESSO \citep{Pepe+2021},
EXPRES \citep{Jurgenson+2016},
KPF \citep{Gibson+2020},
and the planned
HARPS3 \citep{Thompson+2016}
and G-CLEF \citep{Szentgyorgyi+2018},
it might be possible to detect
Earth-mass planets with orbital periods approaching
a year. In addition to finding
potentially habitable planets,
the results from these facilities will provide more overlap between Doppler-based occurrence measurements as a function of mass and transit-based occurrence measurements as a function of radius. This combination will improve our
understanding of 
the compositions of small planets, the
eccentricities of their orbits, and the
mutual inclinations between their orbital planes.

Meanwhile, the PLATO mission
\citet{Rauer+2016}]
is scheduled for launch in 2026 by the European Space Agency. PLATO will
perform a transit survey using $26\times 0.2$\,m optical telescopes with
an combined instantaneous field of view of about 2{,}300 square degrees.
The current plan is to monitor
two fields for two years each, with a top-level goal of finding $\sim$10 habitable-zone Earth-sized planets around Sun-like stars.
A China-based collaboration is proposing
a space-based transit survey
called Earth 2.0 \citep{Ge+2022}
that would
monitor a field that encompasses the original Kepler field.

In the years to come, the domains of all the planet detection
techniques --- including astrometry, gravitational microlensing, and direct imaging --- will begin overlapping.  Some efforts have already
been made to determine occurrence rate densities based on data from
different techniques \citep[see,
  e.g.][]{Montet+2014,ClantonGaudi2016}.
We can look forward to a
more holistic view of the occurrence of planets around other stars,
barring any civilization-ending cataclysm.

\bibliographystyle{spbasicHBexo}  %for bibtex
\bibliography{references} %for bibtex-example

\begin{thebibliography}{134}
\providecommand{\natexlab}[1]{#1}
\providecommand{\url}[1]{{#1}}
\providecommand{\urlprefix}{URL }
\expandafter\ifx\csname urlstyle\endcsname\relax
  \providecommand{\doi}[1]{DOI~\discretionary{}{}{}#1}\else
  \providecommand{\doi}{DOI~\discretionary{}{}{}\begingroup \urlstyle{rm}\Url}\fi
\providecommand{\eprint}[2][]{\url{#2}}

\bibitem[{{Agol} et~al.(2021){Agol}, {Dorn}, {Grimm}, {Turbet}, {Ducrot}, {Delrez}, {Gillon}, {Demory}, {Burdanov}, {Barkaoui}, {Benkhaldoun}, {Bolmont}, {Burgasser}, {Carey}, {de Wit}, {Fabrycky}, {Foreman-Mackey}, {Haldemann}, {Hernandez}, {Ingalls}, {Jehin}, {Langford}, {Leconte}, {Lederer}, {Luger}, {Malhotra}, {Meadows}, {Morris}, {Pozuelos}, {Queloz}, {Raymond}, {Selsis}, {Sestovic}, {Triaud}, and {Van Grootel}}]{Agol+2021}
{Agol} E, {Dorn} C, {Grimm} SL et~al. (2021) {Refining the Transit-timing and Photometric Analysis of TRAPPIST-1: Masses, Radii, Densities, Dynamics, and Ephemerides}. Planetary Science Journal 2(1):1

\bibitem[{{Albrecht} et~al.(2022){Albrecht}, {Dawson}, and {Winn}}]{Albrecht+2022}
{Albrecht} SH, {Dawson} RI {Winn} JN (2022) {Stellar Obliquities in Exoplanetary Systems}. \pasp 134(1038):082001

\bibitem[{{Anglada-Escud{\'e}} et~al.(2016){Anglada-Escud{\'e}}, {Amado}, {Barnes}, {Berdi{\~n}as}, {Butler}, {Coleman}, {de La Cueva}, {Dreizler}, {Endl}, {Giesers}, {Jeffers}, {Jenkins}, {Jones}, {Kiraga}, {K{\"u}rster}, {L{\'o}pez-Gonz{\'a}lez}, {Marvin}, {Morales}, {Morin}, {Nelson}, {Ortiz}, {Ofir}, {Paardekooper}, {Reiners}, {Rodr{\'{\i}}guez}, {Rodr{\'{\i}}guez-L{\'o}pez}, {Sarmiento}, {Strachan}, {Tsapras}, {Tuomi}, and {Zechmeister}}]{AngladaEscude+2016}
{Anglada-Escud{\'e}} G, {Amado} PJ, {Barnes} J et~al. (2016) {A terrestrial planet candidate in a temperate orbit around Proxima Centauri}. \nat 536:437--440

\bibitem[{{Armstrong} et~al.(2014){Armstrong}, {Osborn}, {Brown}, {Faedi}, {G{\'o}mez Maqueo Chew}, {Martin}, {Pollacco}, and {Udry}}]{Armstrong+2014}
{Armstrong} DJ, {Osborn} HP, {Brown} DJA et~al. (2014) {On the abundance of circumbinary planets}. \mnras 444:1873--1883

\bibitem[{{Ballard} and {Johnson}(2016)}]{BallardJohnson2016}
{Ballard} S {Johnson} JA (2016) {The Kepler Dichotomy among the M Dwarfs: Half of Systems Contain Five or More Coplanar Planets}. \apj 816:66

\bibitem[{{Behrens} et~al.(2020){Behrens}, {Ransom}, {Madison}, {Arzoumanian}, {Crowter}, {DeCesar}, {Demorest}, {Dolch}, {Ellis}, {Ferdman}, {Ferrara}, {Fonseca}, {Gentile}, {Jones}, {Jones}, {Lam}, {Levin}, {Lorimer}, {Lynch}, {McLaughlin}, {Ng}, {Nice}, {Pennucci}, {Perera}, {Ray}, {Spiewak}, {Stairs}, {Stovall}, {Swiggum}, and {Zhu}}]{Behrens+2020}
{Behrens} EA, {Ransom} SM, {Madison} DR et~al. (2020) {The NANOGrav 11 yr Data Set: Constraints on Planetary Masses Around 45 Millisecond Pulsars}. \apjl 893(1):L8

\bibitem[{{Beleznay} and {Kunimoto}(2022)}]{Beleznay+2022}
{Beleznay} M {Kunimoto} M (2022) {Exploring the dependence of hot Jupiter occurrence rates on stellar mass with TESS}. \mnras 516(1):75--83

\bibitem[{{Bonfils} et~al.(2013){Bonfils}, {Delfosse}, {Udry}, {Forveille}, {Mayor}, {Perrier}, {Bouchy}, {Gillon}, {Lovis}, {Pepe}, {Queloz}, {Santos}, {S{\'e}gransan}, and {Bertaux}}]{Bonfils+2013}
{Bonfils} X, {Delfosse} X, {Udry} S et~al. (2013) {The HARPS search for southern extra-solar planets. XXXI. The M-dwarf sample}. \aap 549:A109

\bibitem[{{Bonfils} et~al.(2018){Bonfils}, {Astudillo-Defru}, {D{\'\i}az}, {Almenara}, {Forveille}, {Bouchy}, {Delfosse}, {Lovis}, {Mayor}, {Murgas}, {Pepe}, {Santos}, {S{\'e}gransan}, {Udry}, and {W{\"u}nsche}}]{Bonfils+2018}
{Bonfils} X, {Astudillo-Defru} N, {D{\'\i}az} R et~al. (2018) {A temperate exo-Earth around a quiet M dwarf at 3.4 parsec}. \aap 613:A25

\bibitem[{{Bonomo} et~al.(2023){Bonomo}, {Dumusque}, {Massa}, {Mortier}, {Bongiolatti}, {Malavolta}, {Sozzetti}, {Buchhave}, {Damasso}, {Haywood}, {Morbidelli}, {Latham}, {Molinari}, {Pepe}, {Poretti}, {Udry}, {Affer}, {Boschin}, {Charbonneau}, {Cosentino}, {Cretignier}, {Ghedina}, {Lega}, {L{\'o}pez-Morales}, {Margini}, {Mart{\'\i}nez Fiorenzano}, {Mayor}, {Micela}, {Pedani}, {Pinamonti}, {Rice}, {Sasselov}, {Tronsgaard}, and {Vanderburg}}]{Bonomo+2023}
{Bonomo} AS, {Dumusque} X, {Massa} A et~al. (2023) {Cold Jupiters and improved masses in 38 Kepler and K2 small planet systems from 3661 HARPS-N radial velocities. No excess of cold Jupiters in small planet systems}. \aap 677:A33

\bibitem[{{Borucki}(2016)}]{Borucki2016}
{Borucki} WJ (2016) {KEPLER Mission: development and overview}. Reports on Progress in Physics 79(3):036901

\bibitem[{{Bouma} et~al.(2018){Bouma}, {Masuda}, and {Winn}}]{Bouma+2018}
{Bouma} LG, {Masuda} K {Winn} JN (2018) {Biases in Planet Occurrence Caused by Unresolved Binaries in Transit Surveys}. \aj 155(6):244

\bibitem[{{Bryan} et~al.(2016){Bryan}, {Knutson}, {Howard}, {Ngo}, {Batygin}, {Crepp}, {Fulton}, {Hinkley}, {Isaacson}, {Johnson}, {Marcy}, and {Wright}}]{Bryan+2016}
{Bryan} ML, {Knutson} HA, {Howard} AW et~al. (2016) {Statistics of Long Period Gas Giant Planets in Known Planetary Systems}. \apj 821:89

\bibitem[{{Bryan} et~al.(2019){Bryan}, {Knutson}, {Lee}, {Fulton}, {Batygin}, {Ngo}, and {Meshkat}}]{Bryan+2019}
{Bryan} ML, {Knutson} HA, {Lee} EJ et~al. (2019) {An Excess of Jupiter Analogs in Super-Earth Systems}. \aj 157(2):52

\bibitem[{{Bryant} et~al.(2023){Bryant}, {Bayliss}, and {Van Eylen}}]{Bryant+2023}
{Bryant} EM, {Bayliss} D {Van Eylen} V (2023) {The occurrence rate of giant planets orbiting low-mass stars with TESS}. \mnras 521(3):3663--3681

\bibitem[{{Bryson} et~al.(2021){Bryson}, {Kunimoto}, {Kopparapu}, {Coughlin}, {Borucki}, {Koch}, {Aguirre}, {Allen}, {Barentsen}, {Batalha}, {Berger}, {Boss}, {Buchhave}, {Burke}, {Caldwell}, {Campbell}, {Catanzarite}, {Chandrasekaran}, {Chaplin}, {Christiansen}, {Christensen-Dalsgaard}, {Ciardi}, {Clarke}, {Cochran}, {Dotson}, {Doyle}, {Duarte}, {Dunham}, {Dupree}, {Endl}, {Fanson}, {Ford}, {Fujieh}, {Gautier}, {Geary}, {Gilliland}, {Girouard}, {Gould}, {Haas}, {Henze}, {Holman}, {Howard}, {Howell}, {Huber}, {Hunter}, {Jenkins}, {Kjeldsen}, {Kolodziejczak}, {Larson}, {Latham}, {Li}, {Mathur}, {Meibom}, {Middour}, {Morris}, {Morton}, {Mullally}, {Mullally}, {Pletcher}, {Prsa}, {Quinn}, {Quintana}, {Ragozzine}, {Ramirez}, {Sanderfer}, {Sasselov}, {Seader}, {Shabram}, {Shporer}, {Smith}, {Steffen}, {Still}, {Torres}, {Troeltzsch}, {Twicken}, {Uddin}, {Van Cleve}, {Voss}, {Weiss}, {Welsh}, {Wohler}, and {Zamudio}}]{Bryson+2021}
{Bryson} S, {Kunimoto} M, {Kopparapu} RK et~al. (2021) {The Occurrence of Rocky Habitable-zone Planets around Solar-like Stars from Kepler Data}. \aj 161(1):36

\bibitem[{{Buchhave} et~al.(2012){Buchhave}, {Latham}, {Johansen}, {Bizzarro}, {Torres}, {Rowe}, {Batalha}, {Borucki}, {Brugamyer}, {Caldwell}, {Bryson}, {Ciardi}, {Cochran}, {Endl}, {Esquerdo}, {Ford}, {Geary}, {Gilliland}, {Hansen}, {Isaacson}, {Laird}, {Lucas}, {Marcy}, {Morse}, {Robertson}, {Shporer}, {Stefanik}, {Still}, and {Quinn}}]{Buchhave+2012}
{Buchhave} LA, {Latham} DW, {Johansen} A et~al. (2012) {An abundance of small exoplanets around stars with a wide range of metallicities}. \nat 486:375--377

\bibitem[{{Buchhave} et~al.(2018){Buchhave}, {Bitsch}, {Johansen}, {Latham}, {Bizzarro}, {Bieryla}, and {Kipping}}]{Buchhave+2018}
{Buchhave} LA, {Bitsch} B, {Johansen} A et~al. (2018) {Jupiter Analogs Orbit Stars with an Average Metallicity Close to That of the Sun}. \apj 856(1):37

\bibitem[{{Carter} et~al.(2012){Carter}, {Agol}, {Chaplin}, {Basu}, {Bedding}, {Buchhave}, {Christensen-Dalsgaard}, {Deck}, {Elsworth}, {Fabrycky}, {Ford}, {Fortney}, {Hale}, {Handberg}, {Hekker}, {Holman}, {Huber}, {Karoff}, {Kawaler}, {Kjeldsen}, {Lissauer}, {Lopez}, {Lund}, {Lundkvist}, {Metcalfe}, {Miglio}, {Rogers}, {Stello}, {Borucki}, {Bryson}, {Christiansen}, {Cochran}, {Geary}, {Gilliland}, {Haas}, {Hall}, {Howard}, {Jenkins}, {Klaus}, {Koch}, {Latham}, {MacQueen}, {Sasselov}, {Steffen}, {Twicken}, and {Winn}}]{Carter+2012}
{Carter} JA, {Agol} E, {Chaplin} WJ et~al. (2012) {Kepler-36: A Pair of Planets with Neighboring Orbits and Dissimilar Densities}. Science 337(6094):556

\bibitem[{{Chiang} and {Laughlin}(2013)}]{ChiangLaughlin2013}
{Chiang} E {Laughlin} G (2013) {The minimum-mass extrasolar nebula: in situ formation of close-in super-Earths}. \mnras 431:3444--3455

\bibitem[{{Christiansen} et~al.(2023){Christiansen}, {Zink}, {Hardegree-Ullman}, {Fernandes}, {Hopkins}, {Rebull}, {Boley}, {Bergsten}, and {Bhure}}]{Christiansen+2023}
{Christiansen} JL, {Zink} JK, {Hardegree-Ullman} KK et~al. (2023) {Scaling K2. VII. Evidence For a High Occurrence Rate of Hot Sub-Neptunes at Intermediate Ages}. \aj 166(6):248

\bibitem[{{Clanton} and {Gaudi}(2016)}]{ClantonGaudi2016}
{Clanton} C {Gaudi} BS (2016) {Synthesizing Exoplanet Demographics: A Single Population of Long-period Planetary Companions to M Dwarfs Consistent with Microlensing, Radial Velocity, and Direct Imaging Surveys}. \apj 819:125

\bibitem[{{Cumming}(2004)}]{Cumming2004}
{Cumming} A (2004) {Detectability of extrasolar planets in radial velocity surveys}. \mnras 354:1165--1176

\bibitem[{{Cumming} et~al.(2008){Cumming}, {Butler}, {Marcy}, {Vogt}, {Wright}, and {Fischer}}]{Cumming+2008}
{Cumming} A, {Butler} RP, {Marcy} GW et~al. (2008) {The Keck Planet Search: Detectability and the Minimum Mass and Orbital Period Distribution of Extrasolar Planets}. \pasp 120:531

\bibitem[{{Deck} et~al.(2012){Deck}, {Holman}, {Agol}, {Carter}, {Lissauer}, {Ragozzine}, and {Winn}}]{Deck+2012}
{Deck} KM, {Holman} MJ, {Agol} E et~al. (2012) {Rapid Dynamical Chaos in an Exoplanetary System}. \apjl 755:L21

\bibitem[{{Dong} et~al.(2018){Dong}, {Xie}, {Zhou}, {Zheng}, and {Luo}}]{Dong+2018}
{Dong} S, {Xie} JW, {Zhou} JL, {Zheng} Z {Luo} A (2018) {LAMOST telescope reveals that Neptunian cousins of hot Jupiters are mostly single offspring of stars that are rich in heavy elements}. Proceedings of the National Academy of Science 115(2):266--271

\bibitem[{{Dressing} and {Charbonneau}(2015)}]{DressingCharbonneau2015}
{Dressing} CD {Charbonneau} D (2015) {The Occurrence of Potentially Habitable Planets Orbiting M Dwarfs Estimated from the Full Kepler Dataset and an Empirical Measurement of the Detection Sensitivity}. \apj 807:45

\bibitem[{{Endl} et~al.(2006){Endl}, {Cochran}, {K{\"u}rster}, {Paulson}, {Wittenmyer}, {MacQueen}, and {Tull}}]{Endl+2006}
{Endl} M, {Cochran} WD, {K{\"u}rster} M et~al. (2006) {Exploring the Frequency of Close-in Jovian Planets around M Dwarfs}. \apj 649(1):436--443

\bibitem[{{Fabrycky} et~al.(2014){Fabrycky}, {Lissauer}, {Ragozzine}, {Rowe}, {Steffen}, {Agol}, {Barclay}, {Batalha}, {Borucki}, {Ciardi}, {Ford}, {Gautier}, {Geary}, {Holman}, {Jenkins}, {Li}, {Morehead}, {Morris}, {Shporer}, {Smith}, {Still}, and {Van Cleve}}]{Fabrycky+2014}
{Fabrycky} DC, {Lissauer} JJ, {Ragozzine} D et~al. (2014) {Architecture of Kepler's Multi-transiting Systems. II. New Investigations with Twice as Many Candidates}. \apj 790:146

\bibitem[{{Fang} and {Margot}(2013)}]{FangMargot2013}
{Fang} J {Margot} JL (2013) {Are Planetary Systems Filled to Capacity? A Study Based on Kepler Results}. \apj 767:115

\bibitem[{{Fernandes} et~al.(2019){Fernandes}, {Mulders}, {Pascucci}, {Mordasini}, and {Emsenhuber}}]{Fernandes+2019}
{Fernandes} RB, {Mulders} GD, {Pascucci} I, {Mordasini} C {Emsenhuber} A (2019) {Hints for a Turnover at the Snow Line in the Giant Planet Occurrence Rate}. \apj 874(1):81

\bibitem[{{Feynman}(1963)}]{Feynman1963}
{Feynman} RP (1963) {Feynman lectures on physics - Volume 1}

\bibitem[{{Figueira} et~al.(2012){Figueira}, {Marmier}, {Bou{\'e}}, {Lovis}, {Santos}, {Montalto}, {Udry}, {Pepe}, and {Mayor}}]{Figueira+2012}
{Figueira} P, {Marmier} M, {Bou{\'e}} G et~al. (2012) {Comparing HARPS and Kepler surveys. The alignment of multiple-planet systems}. \aap 541:A139

\bibitem[{{Fischer} and {Valenti}(2005)}]{FischerValenti2005}
{Fischer} DA {Valenti} J (2005) {The Planet-Metallicity Correlation}. \apj 622:1102--1117

\bibitem[{{Foreman-Mackey} et~al.(2014){Foreman-Mackey}, {Hogg}, and {Morton}}]{ForemanMackey+2014}
{Foreman-Mackey} D, {Hogg} DW {Morton} TD (2014) {Exoplanet Population Inference and the Abundance of Earth Analogs from Noisy, Incomplete Catalogs}. \apj 795:64

\bibitem[{{Fulton} et~al.(2014){Fulton}, {Tonry}, {Flewelling}, {Burgett}, {Chambers}, {Hodapp}, {Huber}, {Kaiser}, {Wainscoat}, and {Waters}}]{Fulton+2014}
{Fulton} BJ, {Tonry} JL, {Flewelling} H et~al. (2014) {A Search for Planetary Eclipses of White Dwarfs in the Pan-STARRS1 Medium-deep Fields}. \apj 796:114

\bibitem[{{Fulton} et~al.(2017){Fulton}, {Petigura}, {Howard}, {Isaacson}, {Marcy}, {Cargile}, {Hebb}, {Weiss}, {Johnson}, {Morton}, {Sinukoff}, {Crossfield}, and {Hirsch}}]{Fulton+2017}
{Fulton} BJ, {Petigura} EA, {Howard} AW et~al. (2017) {The California-Kepler Survey. III. A Gap in the Radius Distribution of Small Planets}. \aj 154:109

\bibitem[{{Fulton} et~al.(2021){Fulton}, {Rosenthal}, {Hirsch}, {Isaacson}, {Howard}, {Dedrick}, {Sherstyuk}, {Blunt}, {Petigura}, {Knutson}, {Behmard}, {Chontos}, {Crepp}, {Crossfield}, {Dalba}, {Fischer}, {Henry}, {Kane}, {Kosiarek}, {Marcy}, {Rubenzahl}, {Weiss}, and {Wright}}]{Fulton+2021}
{Fulton} BJ, {Rosenthal} LJ, {Hirsch} LA et~al. (2021) {California Legacy Survey. II. Occurrence of Giant Planets beyond the Ice Line}. \apjs 255(1):14

\bibitem[{{Gaidos} et~al.(2016){Gaidos}, {Mann}, {Kraus}, and {Ireland}}]{Gaidos+2016}
{Gaidos} E, {Mann} AW, {Kraus} AL {Ireland} M (2016) {They are small worlds after all: revised properties of Kepler M dwarf stars and their planets}. \mnras 457:2877--2899

\bibitem[{{Gan} et~al.(2023){Gan}, {Wang}, {Wang}, {Mao}, {Huang}, {Collins}, {Stassun}, {Shporer}, {Zhu}, {Ricker}, {Vanderspek}, {Latham}, {Seager}, {Winn}, {Jenkins}, {Barkaoui}, {Belinski}, {Ciardi}, {Evans}, {Girardin}, {Maslennikova}, {Mazeh}, {Panahi}, {Pozuelos}, {Radford}, {Schwarz}, {Twicken}, {W{\"u}nsche}, and {Zucker}}]{Gan+2023}
{Gan} T, {Wang} SX, {Wang} S et~al. (2023) {Occurrence Rate of Hot Jupiters Around Early-type M Dwarfs Based on Transiting Exoplanet Survey Satellite Data}. \aj 165(1):17

\bibitem[{{Ge} et~al.(2022){Ge}, {Zhang}, {Zang}, {Deng}, {Mao}, {Xie}, {Liu}, {Zhou}, {Willis}, {Huang}, {Howell}, {Feng}, {Zhu}, {Yao}, {Liu}, {Aizawa}, {Zhu}, {Li}, {Ma}, {Ye}, {Yu}, {Xiang}, {Yu}, {Liu}, {Yang}, {Wang}, {Shi}, {Fang}, {Zong}, {Liu}, {Zhang}, {Zhang}, {El-Badry}, {Shen}, {Tam}, {Hu}, {Yang}, {Zou}, {Wu}, {Lei}, {Wei}, {Wu}, {Sun}, {Wang}, {Zhang}, {Xu}, {Yang}, {Li}, {Xiang}, {Wang}, {Wang}, {Zhang}, {Jia}, {Yuan}, {Zhang}, {Xuesong Wang}, {Gan}, {Wang}, {Zhao}, {Liu}, {Wei}, {Kang}, {Yang}, {Qi}, {Liu}, {Zhang}, {Zhu}, {Zhou}, {Zhang}, {Yu}, {Zhang}, {Li}, {Tang}, {Wang}, {Wang}, {Li}, {Cheng}, {Shen}, {Li}, {Pan}, {Yang}, {Gao}, {Song}, {Wang}, {Zhang}, {Chen}, {Wang}, {Zhang}, {Wang}, {Zeng}, {Zheng}, {Zhu}, {Guo}, {Zhang}, {Li}, {Wen}, {Feng}, {Chen}, {Chen}, {Han}, {Yang}, {Wang}, {Duan}, {Huang}, {Liang}, {Bi}, {Gai}, {Ge}, {Guo}, {Huang}, {Li}, {Li}, {Li}, {Yuxi}, {Lu}, {Rix}, {Shi}, {Song}, {Tang}, {Ting}, {Wu}, {Wu}, {Yang}, {Yin}, {Gould}, {Lee}, {Dong}, {Yee}, {Shvartzvald},
  {Yang}, {Kuang}, {Zhang}, {Liao}, {Qi}, {Yang}, {Zhang}, {Jiang}, {Ou}, {Li}, {Beck}, {Bedding}, {Campante}, {Chaplin}, {Christensen-Dalsgaard}, {Garc{\'\i}a}, {Gaulme}, {Gizon}, {Hekker}, {Huber}, {Khanna}, {Li}, {Mathur}, {Miglio}, {Mosser}, {Ong}, {Santos}, {Stello}, {Bowman}, {Lares-Martiz}, {Murphy}, {Niu}, {Ma}, {Moln{\'a}r}, {Fu}, {De Cat}, {Su}, and {consortium}}]{Ge+2022}
{Ge} J, {Zhang} H, {Zang} W et~al. (2022) {ET White Paper: To Find the First Earth 2.0}. arXiv e-prints arXiv:2206.06693

\bibitem[{{Gibson} et~al.(2020){Gibson}, {Howard}, {Rider}, {Roy}, {Edelstein}, {Kassis}, {Grillo}, {Halverson}, {Sirk}, {Smith}, {Allen}, {Baker}, {Beichman}, {Berriman}, {Brown}, {Casey}, {Chin}, {Coutts}, {Cowley}, {Deich}, {Feger}, {Fulton}, {Gers}, {Gurevich}, {Ishikawa}, {James}, {Jelinsky}, {Kaye}, {Lanclos}, {Li}, {Lilley}, {McCarney}, {Miller}, {Milner}, {O'Hanlon}, {Pember}, {Raffanti}, {Rockosi}, {Rubenzahl}, {Rumph}, {Sandford}, {Savage}, {Schwab}, {Seifahrt}, {Shaum}, {Smith}, {Stuermer}, {Thorne}, {Vandenberg}, {Von Boeckmann}, {Wang}, {Wang}, {Weisfeiler}, {Wilcox}, {Wishnow}, {Wizinowich}, {Wold}, and {Wolfenberger}}]{Gibson+2020}
{Gibson} SR, {Howard} AW, {Rider} K et~al. (2020) {Keck Planet Finder: design updates}. In: Society of Photo-Optical Instrumentation Engineers (SPIE) Conference Series, Society of Photo-Optical Instrumentation Engineers (SPIE) Conference Series, vol 11447, p 1144742, \doi{10.1117/12.2561783}

\bibitem[{{Gilliland} et~al.(2000){Gilliland}, {Brown}, {Guhathakurta}, {Sarajedini}, {Milone}, {Albrow}, {Baliber}, {Bruntt}, {Burrows}, {Charbonneau}, {Choi}, {Cochran}, {Edmonds}, {Frandsen}, {Howell}, {Lin}, {Marcy}, {Mayor}, {Naef}, {Sigurdsson}, {Stagg}, {Vandenberg}, {Vogt}, and {Williams}}]{Gilliland+2000}
{Gilliland} RL, {Brown} TM, {Guhathakurta} P et~al. (2000) {A Lack of Planets in 47 Tucanae from a Hubble Space Telescope Search}. \apjl 545:L47--L51

\bibitem[{{Ginzburg} et~al.(2018){Ginzburg}, {Schlichting}, and {Sari}}]{Ginzburg+2018}
{Ginzburg} S, {Schlichting} HE {Sari} R (2018) {Core-powered mass-loss and the radius distribution of small exoplanets}. \mnras 476(1):759--765

\bibitem[{{Gonzalez}(1997)}]{Gonzalez+1997}
{Gonzalez} G (1997) {The stellar metallicity-giant planet connection}. \mnras 285(2):403--412

\bibitem[{{Gould} et~al.(2003){Gould}, {Pepper}, and {DePoy}}]{Gould+2003}
{Gould} A, {Pepper} J {DePoy} DL (2003) {Sensitivity of Transit Searches to Habitable-Zone Planets}. \apj 594(1):533--537

\bibitem[{{Grether} and {Lineweaver}(2006)}]{GretherLineweaver2006}
{Grether} D {Lineweaver} CH (2006) {How Dry is the Brown Dwarf Desert? Quantifying the Relative Number of Planets, Brown Dwarfs, and Stellar Companions around Nearby Sun-like Stars}. \apj 640:1051--1062

\bibitem[{{Guo} et~al.(2017){Guo}, {Johnson}, {Mann}, {Kraus}, {Curtis}, and {Latham}}]{Guo+2017}
{Guo} X, {Johnson} JA, {Mann} AW et~al. (2017) {The Metallicity Distribution and Hot Jupiter Rate of the Kepler Field: Hectochelle High-resolution Spectroscopy for 776 Kepler Target Stars}. \apj 838:25

\bibitem[{{Hadden} and {Lithwick}(2014)}]{HaddenLithwick2014}
{Hadden} S {Lithwick} Y (2014) {Densities and Eccentricities of 139 Kepler Planets from Transit Time Variations}. \apj 787:80

\bibitem[{{Hansen} and {Murray}(2012)}]{HansenMurray2012}
{Hansen} BMS {Murray} N (2012) {Migration Then Assembly: Formation of Neptune-mass Planets inside 1 AU}. \apj 751:158

\bibitem[{{He} et~al.(2017){He}, {Triaud}, and {Gillon}}]{He+2017}
{He} MY, {Triaud} AHMJ {Gillon} M (2017) {First limits on the occurrence rate of short-period planets orbiting brown dwarfs}. \mnras 464:2687--2697

\bibitem[{{He} et~al.(2019){He}, {Ford}, and {Ragozzine}}]{He+2019}
{He} MY, {Ford} EB {Ragozzine} D (2019) {Architectures of exoplanetary systems - I. A clustered forward model for exoplanetary systems around Kepler's FGK stars}. \mnras 490(4):4575--4605

\bibitem[{{Hirano} et~al.(2017){Hirano}, {Dai}, {Gandolfi}, {Fukui}, {Livingston}, {Miyakawa}, {Endl}, {Cochran}, {Alonso-Floriano}, {Kuzuhara}, {Montes}, {Ryu}, {Albrecht}, {Barragan}, {Cabrera}, {Csizmadia}, {Deeg}, {Eigm{\"u}ller}, {Erikson}, {Fridlund}, {Grziwa}, {Guenther}, {Hatzes}, {Korth}, {Kudo}, {Kusakabe}, {Narita}, {Nespral}, {Nowak}, {P{\"a}tzold}, {Palle}, {Persson}, {Prieto-Arranz}, {Rauer}, {Ribas}, {Sato}, {Smith}, {Tamura}, {Tanaka}, {Van Eylen}, and {Winn}}]{Hirano+2017}
{Hirano} T, {Dai} F, {Gandolfi} D et~al. (2017) {Planetary Systems around Low-mass Stars Unveiled by K2}. ArXiv e-prints

\bibitem[{{Howard} et~al.(2010){Howard}, {Marcy}, {Johnson}, {Fischer}, {Wright}, {Isaacson}, {Valenti}, {Anderson}, {Lin}, and {Ida}}]{Howard+2010}
{Howard} AW, {Marcy} GW, {Johnson} JA et~al. (2010) {The Occurrence and Mass Distribution of Close-in Super-Earths, Neptunes, and Jupiters}. Science 330:653

\bibitem[{{Howard} et~al.(2012){Howard}, {Marcy}, {Bryson}, {Jenkins}, {Rowe}, {Batalha}, {Borucki}, {Koch}, {Dunham}, {Gautier}, {Van Cleve}, {Cochran}, {Latham}, {Lissauer}, {Torres}, {Brown}, {Gilliland}, {Buchhave}, {Caldwell}, {Christensen-Dalsgaard}, {Ciardi}, {Fressin}, {Haas}, {Howell}, {Kjeldsen}, {Seager}, {Rogers}, {Sasselov}, {Steffen}, {Basri}, {Charbonneau}, {Christiansen}, {Clarke}, {Dupree}, {Fabrycky}, {Fischer}, {Ford}, {Fortney}, {Tarter}, {Girouard}, {Holman}, {Johnson}, {Klaus}, {Machalek}, {Moorhead}, {Morehead}, {Ragozzine}, {Tenenbaum}, {Twicken}, {Quinn}, {Isaacson}, {Shporer}, {Lucas}, {Walkowicz}, {Welsh}, {Boss}, {Devore}, {Gould}, {Smith}, {Morris}, {Prsa}, {Morton}, {Still}, {Thompson}, {Mullally}, {Endl}, and {MacQueen}}]{Howard+2012}
{Howard} AW, {Marcy} GW, {Bryson} ST et~al. (2012) {Planet Occurrence within 0.25 AU of Solar-type Stars from Kepler}. \apjs 201:15

\bibitem[{{Howell} et~al.(2014){Howell}, {Sobeck}, {Haas}, {Still}, {Barclay}, {Mullally}, {Troeltzsch}, {Aigrain}, {Bryson}, {Caldwell}, {Chaplin}, {Cochran}, {Huber}, {Marcy}, {Miglio}, {Najita}, {Smith}, {Twicken}, and {Fortney}}]{Howell+2014}
{Howell} SB, {Sobeck} C, {Haas} M et~al. (2014) {The K2 Mission: Characterization and Early Results}. \pasp 126(938):398

\bibitem[{{Hsu} et~al.(2018){Hsu}, {Ford}, {Ragozzine}, and {Morehead}}]{Hsu+2018}
{Hsu} DC, {Ford} EB, {Ragozzine} D {Morehead} RC (2018) {Improving the Accuracy of Planet Occurrence Rates from Kepler Using Approximate Bayesian Computation}. \aj 155(5):205

\bibitem[{{Huang} et~al.(2016){Huang}, {Wu}, and {Triaud}}]{Huang+2016}
{Huang} C, {Wu} Y {Triaud} AHMJ (2016) {Warm Jupiters Are Less Lonely than Hot Jupiters: Close Neighbors}. \apj 825:98

\bibitem[{{Ida} and {Lin}(2008)}]{IdaLin2008}
{Ida} S {Lin} DNC (2008) {Toward a Deterministic Model of Planetary Formation. V. Accumulation Near the Ice Line and Super-Earths}. \apj 685:584-595

\bibitem[{{Johnson} et~al.(2010{\natexlab{a}}){Johnson}, {Howard}, {Bowler}, {Henry}, {Marcy}, {Wright}, {Fischer}, and {Isaacson}}]{Johnson+2010}
{Johnson} JA, {Howard} AW, {Bowler} BP et~al. (2010{\natexlab{a}}) {Retired A Stars and Their Companions. IV. Seven Jovian Exoplanets from Keck Observatory}. \pasp 122:701

\bibitem[{{Johnson} et~al.(2010{\natexlab{b}}){Johnson}, {Howard}, {Marcy}, {Bowler}, {Henry}, {Fischer}, {Apps}, {Isaacson}, and {Wright}}]{Johnson+2010b}
{Johnson} JA, {Howard} AW, {Marcy} GW et~al. (2010{\natexlab{b}}) {The California Planet Survey. II. A Saturn-Mass Planet Orbiting the M Dwarf Gl 649}. \pasp 122(888):149

\bibitem[{{Jurgenson} et~al.(2016){Jurgenson}, {Fischer}, {McCracken}, {Sawyer}, {Szymkowiak}, {Davis}, {Muller}, and {Santoro}}]{Jurgenson+2016}
{Jurgenson} C, {Fischer} D, {McCracken} T et~al. (2016) {EXPRES: a next generation RV spectrograph in the search for earth-like worlds}. In: {Evans} CJ, {Simard} L {Takami} H (eds) Ground-based and Airborne Instrumentation for Astronomy VI, Society of Photo-Optical Instrumentation Engineers (SPIE) Conference Series, vol 9908, p 99086T, \doi{10.1117/12.2233002}

\bibitem[{{Kasting} et~al.(2014){Kasting}, {Kopparapu}, {Ramirez}, and {Harman}}]{Kasting+2014}
{Kasting} JF, {Kopparapu} R, {Ramirez} RM {Harman} CE (2014) {Remote life-detection criteria, habitable zone boundaries, and the frequency of Earth-like planets around M and late K stars}. Proceedings of the National Academy of Science 111:12,641--12,646

\bibitem[{{Kerr} et~al.(2015){Kerr}, {Johnston}, {Hobbs}, and {Shannon}}]{Kerr+2015}
{Kerr} M, {Johnston} S, {Hobbs} G {Shannon} RM (2015) {Limits on Planet Formation Around Young Pulsars and Implications for Supernova Fallback Disks}. \apjl 809(1):L11

\bibitem[{{Kipping}(2013)}]{Kipping+2013}
{Kipping} DM (2013) {Parametrizing the exoplanet eccentricity distribution with the beta distribution.} \mnras 434:L51--L55

\bibitem[{{Kopparapu} et~al.(2014){Kopparapu}, {Ramirez}, {SchottelKotte}, {Kasting}, {Domagal-Goldman}, and {Eymet}}]{Kopparapu+2014}
{Kopparapu} RK, {Ramirez} RM, {SchottelKotte} J et~al. (2014) {Habitable Zones around Main-sequence Stars: Dependence on Planetary Mass}. \apjl 787:L29

\bibitem[{{Lecar} et~al.(2006){Lecar}, {Podolak}, {Sasselov}, and {Chiang}}]{Lecar+2006}
{Lecar} M, {Podolak} M, {Sasselov} D {Chiang} E (2006) {On the Location of the Snow Line in a Protoplanetary Disk}. \apj 640:1115--1118

\bibitem[{{Lissauer} et~al.(2023){Lissauer}, {Rowe}, {Jontof-Hutter}, {Fabrycky}, {Ford}, {Ragozzine}, {Steffen}, and {Nizam}}]{Lissauer+2023}
{Lissauer} JJ, {Rowe} JF, {Jontof-Hutter} D et~al. (2023) {Updated Catalog of Kepler Planet Candidates: Focus on Accuracy and Orbital Periods}. arXiv e-prints arXiv:2311.00238

\bibitem[{{Lithwick} et~al.(2012){Lithwick}, {Xie}, and {Wu}}]{Lithwick+2012}
{Lithwick} Y, {Xie} J {Wu} Y (2012) {Extracting Planet Mass and Eccentricity from TTV Data}. \apj 761(2):122

\bibitem[{{Lopez} and {Fortney}(2013)}]{LopezFortney2013}
{Lopez} ED {Fortney} JJ (2013) {The Role of Core Mass in Controlling Evaporation: The Kepler Radius Distribution and the Kepler-36 Density Dichotomy}. \apj 776:2

\bibitem[{{Lovis} and {Fischer}(2010)}]{LovisFischer2010}
{Lovis} C {Fischer} D (2010) {Radial Velocity Techniques for Exoplanets}, pp 27--53

\bibitem[{{Luque} and {Pall{\'e}}(2022)}]{LuquePalle2022}
{Luque} R {Pall{\'e}} E (2022) {Density, not radius, separates rocky and water-rich small planets orbiting M dwarf stars}. Science 377(6611):1211--1214

\bibitem[{{Mahadevan} et~al.(2012){Mahadevan}, {Ramsey}, {Bender}, {Terrien}, {Wright}, {Halverson}, {Hearty}, {Nelson}, {Burton}, {Redman}, {Osterman}, {Diddams}, {Kasting}, {Endl}, and {Deshpande}}]{Mahadevan+2012}
{Mahadevan} S, {Ramsey} L, {Bender} C et~al. (2012) {The habitable-zone planet finder: a stabilized fiber-fed NIR spectrograph for the Hobby-Eberly Telescope}. In: {McLean} IS, {Ramsay} SK {Takami} H (eds) Ground-based and Airborne Instrumentation for Astronomy IV, Society of Photo-Optical Instrumentation Engineers (SPIE) Conference Series, vol 8446, p 84461S, \doi{10.1117/12.926102}

\bibitem[{{Mann} et~al.(2017){Mann}, {Gaidos}, {Vanderburg}, {Rizzuto}, {Ansdell}, {Medina}, {Mace}, {Kraus}, and {Sokal}}]{Mann+2017}
{Mann} AW, {Gaidos} E, {Vanderburg} A et~al. (2017) {Zodiacal Exoplanets in Time (ZEIT). IV. Seven Transiting Planets in the Praesepe Cluster}. \aj 153:64

\bibitem[{{Masuda} and {Winn}(2017)}]{MasudaWinn2017}
{Masuda} K {Winn} JN (2017) {Reassessment of the Null Result of the HST Search for Planets in 47 Tucanae}. \aj 153:187

\bibitem[{{Mayor} et~al.(2011){Mayor}, {Marmier}, {Lovis}, {Udry}, {S{\'e}gransan}, {Pepe}, {Benz}, {Bertaux}, {Bouchy}, {Dumusque}, {Lo Curto}, {Mordasini}, {Queloz}, and {Santos}}]{Mayor+2011}
{Mayor} M, {Marmier} M, {Lovis} C et~al. (2011) {The HARPS search for southern extra-solar planets XXXIV. Occurrence, mass distribution and orbital properties of super-Earths and Neptune-mass planets}. arxiv:11092497

\bibitem[{{Mazeh} et~al.(2016){Mazeh}, {Holczer}, and {Faigler}}]{Mazeh+2016}
{Mazeh} T, {Holczer} T {Faigler} S (2016) {Dearth of short-period Neptunian exoplanets: A desert in period-mass and period-radius planes}. \aap 589:A75

\bibitem[{{Millholland} and {Winn}(2021)}]{MillhollandWinn2021}
{Millholland} SC {Winn} JN (2021) {Split Peas in a Pod: Intra-system Uniformity of Super-Earths and Sub-Neptunes}. \apjl 920(2):L34

\bibitem[{{Moe} and {Kratter}(2021)}]{MoeKratter2021}
{Moe} M {Kratter} KM (2021) {Impact of binary stars on planet statistics - I. Planet occurrence rates and trends with stellar mass}. \mnras 507(3):3593--3611

\bibitem[{{Montet} et~al.(2014){Montet}, {Crepp}, {Johnson}, {Howard}, and {Marcy}}]{Montet+2014}
{Montet} BT, {Crepp} JR, {Johnson} JA, {Howard} AW {Marcy} GW (2014) {The TRENDS High-contrast Imaging Survey. IV. The Occurrence Rate of Giant Planets around M Dwarfs}. \apj 781:28

\bibitem[{{Muirhead} et~al.(2015){Muirhead}, {Mann}, {Vanderburg}, {Morton}, {Kraus}, {Ireland}, {Swift}, {Feiden}, {Gaidos}, and {Gazak}}]{Muirhead+2015}
{Muirhead} PS, {Mann} AW, {Vanderburg} A et~al. (2015) {Kepler-445, Kepler-446 and the Occurrence of Compact Multiples Orbiting Mid-M Dwarf Stars}. \apj 801:18

\bibitem[{{Mulders} et~al.(2015){Mulders}, {Pascucci}, and {Apai}}]{Mulders+2015}
{Mulders} GD, {Pascucci} I {Apai} D (2015) {An Increase in the Mass of Planetary Systems around Lower-mass Stars}. \apj 814:130

\bibitem[{{Mulders} et~al.(2016){Mulders}, {Pascucci}, {Apai}, {Frasca}, and {Molenda-Zakowicz}}]{Mulders+2016}
{Mulders} GD, {Pascucci} I, {Apai} D, {Frasca} A {Molenda-Zakowicz} J (2016) {A Super-solar Metallicity for Stars with Hot Rocky Exoplanets}. \aj 152:187

\bibitem[{{Ni{\c{t}}u} et~al.(2022){Ni{\c{t}}u}, {Keith}, {Stappers}, {Lyne}, and {Mickaliger}}]{Nictu+2022}
{Ni{\c{t}}u} IC, {Keith} MJ, {Stappers} BW, {Lyne} AG {Mickaliger} MB (2022) {A search for planetary companions around 800 pulsars from the Jodrell Bank pulsar timing programme}. \mnras 512(2):2446--2459

\bibitem[{{Nutzman} and {Charbonneau}(2008)}]{NutzmanCharbonneau2008}
{Nutzman} P {Charbonneau} D (2008) {Design Considerations for a Ground-Based Transit Search for Habitable Planets Orbiting M Dwarfs}. \pasp 120(865):317

\bibitem[{{Owen} and {Wu}(2013)}]{OwenWu2013}
{Owen} JE {Wu} Y (2013) {Kepler Planets: A Tale of Evaporation}. \apj 775:105

\bibitem[{{Pepe} et~al.(2021){Pepe}, {Cristiani}, {Rebolo}, {Santos}, {Dekker}, {Cabral}, {Di Marcantonio}, {Figueira}, {Lo Curto}, {Lovis}, {Mayor}, {M{\'e}gevand}, {Molaro}, {Riva}, {Zapatero Osorio}, {Amate}, {Manescau}, {Pasquini}, {Zerbi}, {Adibekyan}, {Abreu}, {Affolter}, {Alibert}, {Aliverti}, {Allart}, {Allende Prieto}, {{\'A}lvarez}, {Alves}, {Avila}, {Baldini}, {Bandy}, {Barros}, {Benz}, {Bianco}, {Borsa}, {Bourrier}, {Bouchy}, {Broeg}, {Calderone}, {Cirami}, {Coelho}, {Conconi}, {Coretti}, {Cumani}, {Cupani}, {D'Odorico}, {Damasso}, {Deiries}, {Delabre}, {Demangeon}, {Dumusque}, {Ehrenreich}, {Faria}, {Fragoso}, {Genolet}, {Genoni}, {G{\'e}nova Santos}, {Gonz{\'a}lez Hern{\'a}ndez}, {Hughes}, {Iwert}, {Kerber}, {Knudstrup}, {Landoni}, {Lavie}, {Lillo-Box}, {Lizon}, {Maire}, {Martins}, {Mehner}, {Micela}, {Modigliani}, {Monteiro}, {Monteiro}, {Moschetti}, {Murphy}, {Nunes}, {Oggioni}, {Oliveira}, {Oshagh}, {Pall{\'e}}, {Pariani}, {Poretti}, {Rasilla}, {Rebord{\~a}o}, {Redaelli}, {Santana Tschudi},
  {Santin}, {Santos}, {S{\'e}gransan}, {Schmidt}, {Segovia}, {Sosnowska}, {Sozzetti}, {Sousa}, {Span{\`o}}, {Su{\'a}rez Mascare{\~n}o}, {Tabernero}, {Tenegi}, {Udry}, and {Zanutta}}]{Pepe+2021}
{Pepe} F, {Cristiani} S, {Rebolo} R et~al. (2021) {ESPRESSO at VLT. On-sky performance and first results}. \aap 645:A96

\bibitem[{{Pepper} et~al.(2003){Pepper}, {Gould}, and {Depoy}}]{Pepper+2003}
{Pepper} J, {Gould} A {Depoy} DL (2003) {Using All-Sky Surveys to Find Planetary Transits}. \actaa 53:213--228

\bibitem[{{Petigura} et~al.(2018){Petigura}, {Marcy}, {Winn}, {Weiss}, {Fulton}, {Howard}, {Sinukoff}, {Isaacson}, {Morton}, and {Johnson}}]{Petigura+2018}
{Petigura} EA, {Marcy} GW, {Winn} JN et~al. (2018) {The California-Kepler Survey. IV. Metal-rich Stars Host a Greater Diversity of Planets}. \aj 155:89

\bibitem[{{Pollack} et~al.(1996){Pollack}, {Hubickyj}, {Bodenheimer}, {Lissauer}, {Podolak}, and {Greenzweig}}]{Pollack+1996}
{Pollack} JB, {Hubickyj} O, {Bodenheimer} P et~al. (1996) {Formation of the Giant Planets by Concurrent Accretion of Solids and Gas}. \icarus 124:62--85

\bibitem[{{Pu} and {Wu}(2015)}]{PuWu2015}
{Pu} B {Wu} Y (2015) {Spacing of Kepler Planets: Sculpting by Dynamical Instability}. \apj 807:44

\bibitem[{{Rasio} and {Ford}(1996)}]{RasioFord1996}
{Rasio} FA {Ford} EB (1996) {Dynamical instabilities and the formation of extrasolar planetary systems}. Science 274:954--956

\bibitem[{{Rauer} et~al.(2016){Rauer}, {Aerts}, {Cabrera}, and {PLATO Team}}]{Rauer+2016}
{Rauer} H, {Aerts} C, {Cabrera} J {PLATO Team} (2016) {The PLATO Mission}. Astronomische Nachrichten 337(8-9):961

\bibitem[{{Reffert} et~al.(2015){Reffert}, {Bergmann}, {Quirrenbach}, {Trifonov}, and {K{\"u}nstler}}]{Reffert+2015}
{Reffert} S, {Bergmann} C, {Quirrenbach} A, {Trifonov} T {K{\"u}nstler} A (2015) {Precise radial velocities of giant stars. VII. Occurrence rate of giant extrasolar planets as a function of mass and metallicity}. \aap 574:A116

\bibitem[{{Ricker} et~al.(2015){Ricker}, {Winn}, {Vanderspek}, {Latham}, {Bakos}, {Bean}, {Berta-Thompson}, {Brown}, {Buchhave}, {Butler}, {Butler}, {Chaplin}, {Charbonneau}, {Christensen-Dalsgaard}, {Clampin}, {Deming}, {Doty}, {De Lee}, {Dressing}, {Dunham}, {Endl}, {Fressin}, {Ge}, {Henning}, {Holman}, {Howard}, {Ida}, {Jenkins}, {Jernigan}, {Johnson}, {Kaltenegger}, {Kawai}, {Kjeldsen}, {Laughlin}, {Levine}, {Lin}, {Lissauer}, {MacQueen}, {Marcy}, {McCullough}, {Morton}, {Narita}, {Paegert}, {Palle}, {Pepe}, {Pepper}, {Quirrenbach}, {Rinehart}, {Sasselov}, {Sato}, {Seager}, {Sozzetti}, {Stassun}, {Sullivan}, {Szentgyorgyi}, {Torres}, {Udry}, and {Villasenor}}]{Ricker+2015}
{Ricker} GR, {Winn} JN, {Vanderspek} R et~al. (2015) {Transiting Exoplanet Survey Satellite (TESS)}. Journal of Astronomical Telescopes, Instruments, and Systems 1(1):014003

\bibitem[{{Rogers} et~al.(2023){Rogers}, {Schlichting}, and {Owen}}]{Rogers+2023}
{Rogers} JG, {Schlichting} HE {Owen} JE (2023) {Conclusive Evidence for a Population of Water Worlds around M Dwarfs Remains Elusive}. \apjl 947(1):L19

\bibitem[{{Rosenthal} et~al.(2021){Rosenthal}, {Fulton}, {Hirsch}, {Isaacson}, {Howard}, {Dedrick}, {Sherstyuk}, {Blunt}, {Petigura}, {Knutson}, {Behmard}, {Chontos}, {Crepp}, {Crossfield}, {Dalba}, {Fischer}, {Henry}, {Kane}, {Kosiarek}, {Marcy}, {Rubenzahl}, {Weiss}, and {Wright}}]{Rosenthal+2021}
{Rosenthal} LJ, {Fulton} BJ, {Hirsch} LA et~al. (2021) {The California Legacy Survey. I. A Catalog of 178 Planets from Precision Radial Velocity Monitoring of 719 Nearby Stars over Three Decades}. \apjs 255(1):8

\bibitem[{{Rosenthal} et~al.(2022){Rosenthal}, {Knutson}, {Chachan}, {Dai}, {Howard}, {Fulton}, {Chontos}, {Crepp}, {Dalba}, {Henry}, {Kane}, {Petigura}, {Weiss}, and {Wright}}]{Rosenthal+2022}
{Rosenthal} LJ, {Knutson} HA, {Chachan} Y et~al. (2022) {The California Legacy Survey. III. On the Shoulders of (Some) Giants: The Relationship between Inner Small Planets and Outer Massive Planets}. \apjs 262(1):1

\bibitem[{{Sabotta} et~al.(2021){Sabotta}, {Schlecker}, {Chaturvedi}, {Guenther}, {Mu{\~n}oz Rodr{\'\i}guez}, {Mu{\~n}oz S{\'a}nchez}, {Caballero}, {Shan}, {Reffert}, {Ribas}, {Reiners}, {Hatzes}, {Amado}, {Klahr}, {Morales}, {Quirrenbach}, {Henning}, {Dreizler}, {Pall{\'e}}, {Perger}, {Azzaro}, {Jeffers}, {Kaminski}, {K{\"u}rster}, {Lafarga}, {Montes}, {Passegger}, and {Zechmeister}}]{Sabotta+2021}
{Sabotta} S, {Schlecker} M, {Chaturvedi} P et~al. (2021) {The CARMENES search for exoplanets around M dwarfs. Planet occurrence rates from a subsample of 71 stars}. \aap 653:A114

\bibitem[{{Sahlmann} et~al.(2011){Sahlmann}, {S{\'e}gransan}, {Queloz}, {Udry}, {Santos}, {Marmier}, {Mayor}, {Naef}, {Pepe}, and {Zucker}}]{Sahlmann+2011}
{Sahlmann} J, {S{\'e}gransan} D, {Queloz} D et~al. (2011) {Search for brown-dwarf companions of stars}. \aap 525:A95

\bibitem[{{Sanchis-Ojeda} et~al.(2014){Sanchis-Ojeda}, {Rappaport}, {Winn}, {Kotson}, {Levine}, and {El Mellah}}]{SanchisOjeda+2014}
{Sanchis-Ojeda} R, {Rappaport} S, {Winn} JN et~al. (2014) {A Study of the Shortest-period Planets Found with Kepler}. \apj 787:47

\bibitem[{{Santerne} et~al.(2016){Santerne}, {Moutou}, {Tsantaki}, {Bouchy}, {H{\'e}brard}, {Adibekyan}, {Almenara}, {Amard}, {Barros}, {Boisse}, {Bonomo}, {Bruno}, {Courcol}, {Deleuil}, {Demangeon}, {D{\'{\i}}az}, {Guillot}, {Havel}, {Montagnier}, {Rajpurohit}, {Rey}, and {Santos}}]{Santerne+2016}
{Santerne} A, {Moutou} C, {Tsantaki} M et~al. (2016) {SOPHIE velocimetry of Kepler transit candidates. XVII. The physical properties of giant exoplanets within 400 days of period}. \aap 587:A64

\bibitem[{{Santos} et~al.(2003){Santos}, {Israelian}, {Mayor}, {Rebolo}, and {Udry}}]{Santos+2003}
{Santos} NC, {Israelian} G, {Mayor} M, {Rebolo} R {Udry} S (2003) {Statistical properties of exoplanets. II. Metallicity, orbital parameters, and space velocities}. \aap 398:363--376

\bibitem[{{Santos} et~al.(2017){Santos}, {Adibekyan}, {Figueira}, {Andreasen}, {Barros}, {Delgado-Mena}, {Demangeon}, {Faria}, {Oshagh}, {Sousa}, {Viana}, and {Ferreira}}]{Santos+2017}
{Santos} NC, {Adibekyan} V, {Figueira} P et~al. (2017) {Observational evidence for two distinct giant planet populations}. \aap 603:A30

\bibitem[{{Schlaufman}(2018)}]{Schlaufman2018}
{Schlaufman} KC (2018) {Evidence of an Upper Bound on the Masses of Planets and Its Implications for Giant Planet Formation}. \apj 853:37

\bibitem[{{Schlaufman} and {Winn}(2016)}]{SchlaufmanWinn2016}
{Schlaufman} KC {Winn} JN (2016) {The Occurrence of Additional Giant Planets Inside the Water-Ice Line in Systems with Hot Jupiters: Evidence Against High-Eccentricity Migration}. \apj 825:62

\bibitem[{{Seifahrt} et~al.(2016){Seifahrt}, {Bean}, {St{\"u}rmer}, {Gers}, {Grobler}, {Reed}, and {Jones}}]{Seifahrt+2016}
{Seifahrt} A, {Bean} JL, {St{\"u}rmer} J et~al. (2016) {Development and construction of MAROON-X}. In: {Evans} CJ, {Simard} L {Takami} H (eds) Ground-based and Airborne Instrumentation for Astronomy VI, Society of Photo-Optical Instrumentation Engineers (SPIE) Conference Series, vol 9908, p 990818, \doi{10.1117/12.2232069}

\bibitem[{{Szab{\'o}} and {Kiss}(2011)}]{SzaboKiss2011}
{Szab{\'o}} GM {Kiss} LL (2011) {A Short-period Censor of Sub-Jupiter Mass Exoplanets with Low Density}. \apjl 727:L44

\bibitem[{{Szentgyorgyi} et~al.(2018){Szentgyorgyi}, {Baldwin}, {Barnes}, {Bean}, {Ben-Ami}, {Brennan}, {Budynkiewicz}, {Catropa}, {Chun}, {Conroy}, {Contos}, {Crane}, {Durusky}, {Epps}, {Evans}, {Evans}, {Fishman}, {Frebel}, {Gauron}, {Guzman}, {Hare}, {Jang}, {Jang}, {Jordan}, {Kim}, {Kim}, {Kim}, {Lee}, {Lopez-Morales}, {Mendes de Oliveira}, {McCracken}, {McMuldroch}, {Miller}, {Mueller}, {Oh}, {Onyuksel}, {Park}, {Park}, {Park}, {Paxson}, {Phillips}, {Plummer}, {Podgorski}, {Rubin}, {Seifahrt}, {Stark}, {Steiner}, {Uomoto}, {Walsworth}, and {Yu}}]{Szentgyorgyi+2018}
{Szentgyorgyi} A, {Baldwin} D, {Barnes} S et~al. (2018) {The GMT-consortium large earth finder (G-CLEF): an optical echelle spectrograph for the Giant Magellan Telescope (GMT)}. In: {Evans} CJ, {Simard} L {Takami} H (eds) Ground-based and Airborne Instrumentation for Astronomy VII, Society of Photo-Optical Instrumentation Engineers (SPIE) Conference Series, vol 10702, p 107021R, \doi{10.1117/12.2313539}

\bibitem[{{Tabachnik} and {Tremaine}(2002)}]{TabachnikTremaine2002}
{Tabachnik} S {Tremaine} S (2002) {Maximum-likelihood method for estimating the mass and period distributions of extrasolar planets}. \mnras 335:151--158

\bibitem[{{Thompson} et~al.(2016){Thompson}, {Queloz}, {Baraffe}, {Brake}, {Dolgopolov}, {Fisher}, {Fleury}, {Geelhoed}, {Hall}, {Gonz{\'a}lez Hern{\'a}ndez}, {ter Horst}, {Kragt}, {Navarro}, {Naylor}, {Pepe}, {Piskunov}, {Rebolo}, {Sander}, {S{\'e}gransan}, {Seneta}, {Sing}, {Snellen}, {Snik}, {Spronck}, {Stempels}, {Sun}, {Santana Tschudi}, and {Young}}]{Thompson+2016}
{Thompson} SJ, {Queloz} D, {Baraffe} I et~al. (2016) {HARPS3 for a roboticized Isaac Newton Telescope}. In: {Evans} CJ, {Simard} L {Takami} H (eds) Ground-based and Airborne Instrumentation for Astronomy VI, Society of Photo-Optical Instrumentation Engineers (SPIE) Conference Series, vol 9908, p 99086F, \doi{10.1117/12.2232111}

\bibitem[{{Tremaine} and {Dong}(2012)}]{TremaineDong2012}
{Tremaine} S {Dong} S (2012) {The Statistics of Multi-planet Systems}. \aj 143:94

\bibitem[{{Triaud} et~al.(2017){Triaud}, {Martin}, {S{\'e}gransan}, {Smalley}, {Maxted}, {Anderson}, {Bouchy}, {Collier Cameron}, {Faedi}, {G{\'o}mez Maqueo Chew}, {Hebb}, {Hellier}, {Marmier}, {Pepe}, {Pollacco}, {Queloz}, {Udry}, and {West}}]{Triaud+2017}
{Triaud} AHMJ, {Martin} DV, {S{\'e}gransan} D et~al. (2017) {The EBLM Project. IV. Spectroscopic orbits of over 100 eclipsing M dwarfs masquerading as transiting hot Jupiters}. \aap 608:A129

\bibitem[{{Van Eylen} and {Albrecht}(2015)}]{VanEylenAlbrecht2015}
{Van Eylen} V {Albrecht} S (2015) {Eccentricity from Transit Photometry: Small Planets in Kepler Multi-planet Systems Have Low Eccentricities}. \apj 808:126

\bibitem[{{Van Eylen} et~al.(2017){Van Eylen}, {Agentoft}, {Lundkvist}, {Kjeldsen}, {Owen}, {Fulton}, {Petigura}, and {Snellen}}]{VanEylen+2017}
{Van Eylen} V, {Agentoft} C, {Lundkvist} MS et~al. (2017) {An asteroseismic view of the radius valley: stripped cores, not born rocky}. ArXiv e-prints

\bibitem[{{van Sluijs} and {Van Eylen}(2018)}]{VanSluijsVanEylen2018}
{van Sluijs} L {Van Eylen} V (2018) {The occurrence of planets and other substellar bodies around white dwarfs using K2}. \mnras 474:4603--4611

\bibitem[{{Weiss} et~al.(2018){Weiss}, {Marcy}, {Petigura}, {Fulton}, {Howard}, {Winn}, {Isaacson}, {Morton}, {Hirsch}, {Sinukoff}, {Cumming}, {Hebb}, and {Cargile}}]{Weiss+2018}
{Weiss} LM, {Marcy} GW, {Petigura} EA et~al. (2018) {The California-Kepler Survey. V. Peas in a Pod: Planets in a Kepler Multi-planet System Are Similar in Size and Regularly Spaced}. \aj 155:48

\bibitem[{{Weiss} et~al.(2023){Weiss}, {Millholland}, {Petigura}, {Adams}, {Batygin}, {Block}, and {Mordasini}}]{Weiss+2023}
{Weiss} LM, {Millholland} SC, {Petigura} EA et~al. (2023) {Architectures of Compact Multi-Planet Systems: Diversity and Uniformity}. In: {Inutsuka} S, {Aikawa} Y, {Muto} T, {Tomida} K {Tamura} M (eds) Astronomical Society of the Pacific Conference Series, Astronomical Society of the Pacific Conference Series, vol 534, p 863

\bibitem[{{Wilson} et~al.(2018){Wilson}, {Teske}, {Majewski}, {Cunha}, {Smith}, {Souto}, {Bender}, {Mahadevan}, {Troup}, {Allende Prieto}, {Stassun}, {Skrutskie}, {Almeida}, {Garc{\'{\i}}a-Hern{\'a}ndez}, {Zamora}, and {Brinkmann}}]{Wilson+2018}
{Wilson} RF, {Teske} J, {Majewski} SR et~al. (2018) {Elemental Abundances of Kepler Objects of Interest in APOGEE. I. Two Distinct Orbital Period Regimes Inferred from Host Star Iron Abundances}. \aj 155:68

\bibitem[{{Winn}(2010)}]{Winn2010}
{Winn} JN (2010) {Exoplanet Transits and Occultations}, University of Arizona Press, pp 55--77

\bibitem[{{Winn} et~al.(2017){Winn}, {Sanchis-Ojeda}, {Rogers}, {Petigura}, {Howard}, {Isaacson}, {Marcy}, {Schlaufman}, {Cargile}, and {Hebb}}]{Winn+2017}
{Winn} JN, {Sanchis-Ojeda} R, {Rogers} L et~al. (2017) {Absence of a Metallicity Effect for Ultra-short-period Planets}. \aj 154(2):60

\bibitem[{{Wittenmyer} et~al.(2020){Wittenmyer}, {Wang}, {Horner}, {Butler}, {Tinney}, {Carter}, {Wright}, {Jones}, {Bailey}, {O'Toole}, and {Johns}}]{Wittenmyer+2020}
{Wittenmyer} RA, {Wang} S, {Horner} J et~al. (2020) {Cool Jupiters greatly outnumber their toasty siblings: occurrence rates from the Anglo-Australian Planet Search}. \mnras 492(1):377--383

\bibitem[{{Wright}(2018)}]{Wright2018}
{Wright} JT (2018) {Radial Velocities as an Exoplanet Discovery Method}. In: {Deeg} HJ {Belmonte} JA (eds) Handbook of Exoplanets, p~4, \doi{10.1007/978-3-319-55333-7_4}

\bibitem[{{Wright} et~al.(2011){Wright}, {Veras}, {Ford}, {Johnson}, {Marcy}, {Howard}, {Isaacson}, {Fischer}, {Spronck}, {Anderson}, and {Valenti}}]{Wright+2011}
{Wright} JT, {Veras} D, {Ford} EB et~al. (2011) {The California Planet Survey. III. A Possible 2:1 Resonance in the Exoplanetary Triple System HD 37124}. \apj 730:93

\bibitem[{{Wright} et~al.(2012){Wright}, {Marcy}, {Howard}, {Johnson}, {Morton}, and {Fischer}}]{Wright+2012}
{Wright} JT, {Marcy} GW, {Howard} AW et~al. (2012) {The Frequency of Hot Jupiters Orbiting nearby Solar-type Stars}. \apj 753:160

\bibitem[{{Xie} et~al.(2016){Xie}, {Dong}, {Zhu}, {Huber}, {Zheng}, {De Cat}, {Fu}, {Liu}, {Luo}, {Wu}, {Zhang}, {Zhang}, {Zhou}, {Cao}, {Hou}, {Wang}, and {Zhang}}]{Xie+2016}
{Xie} JW, {Dong} S, {Zhu} Z et~al. (2016) {Exoplanet orbital eccentricities derived from LAMOST-Kepler analysis}. Proceedings of the National Academy of Science 113:11,431--11,435

\bibitem[{{Youdin}(2011)}]{Youdin2011}
{Youdin} AN (2011) {The Exoplanet Census: A General Method Applied to Kepler}. \apj 742:38

\bibitem[{{Zhou} et~al.(2019){Zhou}, {Huang}, {Bakos}, {Hartman}, {Latham}, {Quinn}, {Collins}, {Winn}, {Wong}, {Kov{\'a}cs}, {Csubry}, {Bhatti}, {Penev}, {Bieryla}, {Esquerdo}, {Berlind}, {Calkins}, {de Val-Borro}, {Noyes}, {L{\'a}z{\'a}r}, {Papp}, {S{\'a}ri}, {Kov{\'a}cs}, {Buchhave}, {Szklenar}, {B{\'e}ky}, {Johnson}, {Cochran}, {Kniazev}, {Stassun}, {Fulton}, {Shporer}, {Espinoza}, {Bayliss}, {Everett}, {Howell}, {Hellier}, {Anderson}, {Collier Cameron}, {West}, {Brown}, {Schanche}, {Barkaoui}, {Pozuelos}, {Gillon}, {Jehin}, {Benkhaldoun}, {Daassou}, {Ricker}, {Vanderspek}, {Seager}, {Jenkins}, {Lissauer}, {Armstrong}, {Collins}, {Gan}, {Hart}, {Horne}, {Kielkopf}, {Nielsen}, {Nishiumi}, {Narita}, {Palle}, {Relles}, {Sefako}, {Tan}, {Davies}, {Goeke}, {Guerrero}, {Haworth}, and {Villanueva}}]{Zhou+2019}
{Zhou} G, {Huang} CX, {Bakos} G{\'A} et~al. (2019) {Two New HATNet Hot Jupiters around A Stars and the First Glimpse at the Occurrence Rate of Hot Jupiters from TESS}. \aj 158(4):141

\bibitem[{{Zhu}(2023)}]{Zhu2023}
{Zhu} W (2023) {The Metallicity Dimension of the Super Earth-Cold Jupiter Correlation}. arXiv e-prints arXiv:2306.16691

\bibitem[{{Zhu} and {Wu}(2018)}]{ZhuWu2018}
{Zhu} W {Wu} Y (2018) {The Super Earth-Cold Jupiter Relations}. \aj 156(3):92

\bibitem[{{Zhu} et~al.(2018){Zhu}, {Petrovich}, {Wu}, {Dong}, and {Xie}}]{Zhu+2018}
{Zhu} W, {Petrovich} C, {Wu} Y, {Dong} S {Xie} J (2018) {About 30\% of Sun-like Stars Have Kepler-like Planetary Systems: A Study of Their Intrinsic Architecture}. \apj 860(2):101

\bibitem[{{Zink} and {Howard}(2023)}]{ZinkHoward2023}
{Zink} JK {Howard} AW (2023) {Hot Jupiters Have Giant Companions: Evidence for Coplanar High-eccentricity Migration}. \apjl 956(1):L29

\bibitem[{{Zink} et~al.(2019){Zink}, {Christiansen}, and {Hansen}}]{Zink+2019}
{Zink} JK, {Christiansen} JL {Hansen} BMS (2019) {Accounting for incompleteness due to transit multiplicity in Kepler planet occurrence rates}. \mnras 483(4):4479--4494

\bibitem[{{Zink} et~al.(2023){Zink}, {Hardegree-Ullman}, {Christiansen}, {Petigura}, {Boley}, {Bhure}, {Rice}, {Yee}, {Isaacson}, {Fernandes}, {Howard}, {Blunt}, {Lubin}, {Chontos}, {Pidhorodetska}, and {MacDougall}}]{Zink+2023}
{Zink} JK, {Hardegree-Ullman} KK, {Christiansen} JL et~al. (2023) {Scaling K2. VI. Reduced Small-planet Occurrence in High-galactic-amplitude Stars}. \aj 165(6):262

\end{thebibliography}

\end{document}